\RequirePackage[l2tabu, orthodox]{nag}
\RequirePackage{snapshot}

\documentclass[9pt,onecolumn]{extarticle}

\makeatletter
\if@twocolumn
  \usepackage[dvips,letterpaper,top=0.5in, bottom=0.5in, left=0.75in, right=0.5in,includefoot,heightrounded]{geometry}
\else
  \usepackage[dvips,letterpaper,margin=1in,includefoot,heightrounded]{geometry}
\fi

\usepackage{srcltx}

\usepackage[russian,portuges,english]{babel}

\iflanguage{portuges}
    {\newcommand{\keywordname}{Palavras-chaves}}
    {\newcommand{\keywordname}{Keywords}}

\usepackage{amsmath}
\usepackage{amssymb,amsfonts}

\usepackage{abstract}

\usepackage{graphicx}
\usepackage[usenames,dvipsnames,svgnames,x11names]{xcolor}

\usepackage[labelformat=simple]{subcaption}

\usepackage{booktabs}

\usepackage{setspace}
\usepackage{flushend}

\usepackage{cite}

\usepackage{hyperref}
\usepackage[normalem]{ulem}

\usepackage{enumerate}

\usepackage{multirow}
\usepackage[noend]{algpseudocode}

\newcommand{\cas}{\operatorname{cas}}
\newcommand{\A}{\alpha}
\newcommand{\B}{\beta}

\usepackage{listings}

\usepackage[short,12hr]{datetime}
       \usepackage{fouriernc}
\makeatletter

\newcommand{\printtitle}{%
\makeatletter
\if@twocolumn

\twocolumn[%
  \maketitle
  \begin{onecolabstract}
    \myabstract
  \end{onecolabstract}
  \begin{center}
    \small
    \textbf{\keywordname}
    \\\medskip
    \mykeywords
  \end{center}
  \bigskip
]
\saythanks
\else
  \maketitle
  \begin{onecolabstract}
    \myabstract
  \end{onecolabstract}
  \begin{center}
    \small
    \textbf{\keywordname}
    \\\medskip
    \mykeywords
  \end{center}
  \bigskip
  \onehalfspacing
\fi
\makeatother
}

\author{%
V.~A.~Coutinho%
\thanks{
Graduate School in Electrical Engineering,
Universidade Federal de Pernambuco (UFPE),
Recife, Brazil.}
\and
F.~M.~Bayer%
\thanks{%
Departamento de Estat\'istica and LACESM,
Universidade Federal de Santa Maria,
Santa Maria, Brazil.}
\and
R.~J.~Cintra%
\thanks{%
Signal Processing Group,
DE/CCEN, UFPE, Recife, Brazil.
E-mail: \url{rjdsc@de.ufpe.br}}
}

\title{%
Low-complexity Three-dimensional Discrete Hartley Transform Approximations for Medical Image Compression}

\newcommand{\myabstract}{%
The discrete Hartley transform~(DHT) is a useful tool for medical image coding.
The three-dimensional DHT~(3D~DHT) can be employed to compress medical image data, such as magnetic resonance and X-ray angiography.
However,
the computation of the 3D~DHT involves several multiplications by irrational quantities,
which require floating-point arithmetic and inherent truncation errors.
In recent years,
a significant progress in wireless and implantable biomedical devices has been achieved.
Such devices present critical power and hardware limitations.
The multiplication operation demands higher hardware, power, and time consumption than other arithmetic operations,
such as addition and bit-shifts.
In this work,
we present a set of multiplierless DHT approximations,
which can be implemented with fixed-point arithmetic.
We derive 3D~DHT approximations by employing tensor formalism.
Such proposed methods present prominent computational savings compared to the
usual 3D~DHT approach,
being appropriate for devices with limited resources.
The proposed transforms are applied in a
lossy
3D~DHT-based medical image
compression algorithm,
presenting practically the same level of visual quality
($>98\%$ in terms of SSIM) at a considerable reduction in computational effort ($100 \%$ multiplicative complexity reduction).
Furthermore,
we implemented the proposed 3D transforms in an
ARM Cortex-M0+ processor employing the low-cost Raspberry Pi Pico board.
The execution time was reduced by $\sim$70 \% compared to the usual 3D~DHT and $\sim$90~\% compared to 3D~DCT.
}

\newcommand{\mykeywords}{%
DHT approximation,
3D~DHT,
DICOM,
data compression,
video coding
}

\date{}

\begin{document}

\printtitle

\section{Introduction}

Discrete
transforms constitute central mathematical apparatuses
in digital signal processing technologies~\cite{ahmed1975}.
The use of
such tools
in real-world applications
demands
fast algorithms~\cite{Blahut2010}
capable of
reducing the
computational overhead
compared to
direct computation.
Fast algorithms have been extensively studied and developed
for several discrete transforms,
such as the discrete Fourier
transform~(DFT)~\cite{duhamel1990fast}
and
the discrete cosine
transform~(DCT)~\cite{Loeffler1989,fw1992}.
Generally,
the computational complexity of fast algorithms
is quantitatively evaluated
by
the number of arithmetic operations~\cite[p.~716]{oppenheim2010discrete}.
It is a well-known fact that
the physical implementation
of
the
multiplication operation
typically requires
a larger amount of hardware and energy resources
relative to
additions and bit-shifting operations~\cite{Blahut2010}.
In this context,
the multiplicative complexity theory
for discrete transforms algorithms was
developed
and
theoretical
lower bounds
for
the multiplicative cost
of
sinusoidal transforms
was advanced in~\cite{Heideman1988}.

The
discrete Hartley transform~(DHT)
was introduced in~\cite{bracewell1983discrete}
and
can be considered as a real-valued
alternative
for the
DFT~\cite{voronenko2009algebraic}.
In fact,
in~\cite[p.~116]{Heideman1988} it was proved that
the DHT and the DFT are equivalent systems
from the point of view of
multiplicative complexity.
Consequently,
both transforms present equal multiplicative complexity lower bounds for the same transform length.
However,
the DHT possesses the advantages of being real-valued and
presenting identical direct and inverse
transformations~\cite{bracewell1983discrete,Gonzalez2001,chiper2013novel}.
Furthermore,
the DHT is also an alternative
to compute other sinusoidal discrete transforms,
such as
the DCT~\cite{rao2014discrete}
and
the discrete sine transform~\cite{meher2006scalable}.
Besides being considered as an auxiliary tool
for computing other discrete transforms,
the DHT has proved to have itself
a range of applications
in diverse areas,
such as
image processing~\cite{narendra2016hartley,dousty2016multifocus},
image compression~\cite{sunder2006medical, jiang2010novel}
image encryption~\cite{liu2010color, bas2013},
watermark image authentication~\cite{mandal2013separable}, and
medical image enhancement~\cite{hossain2010medical}.
A collection of fast algorithms and hardware architectures for the DHT
can be found in literature~\cite{meher2006scalable, grigoryan2004novel,
	guo2000efficient,hou1987fastdht,bracewell1986fast}.

Although
fast algorithms for the aforementioned sinusoidal discrete transforms
generate
a remarkable
reduction in computational costs~\cite{Blahut2010},
they
are still restricted to
the theoretical multiplicative lower bounds~\cite{Heideman1988}.
The significant research effort in the field
of fast algorithms
has led
to methods
capable of
achieving
the minimum multiplicative complexity,
as it is the case
for the
widely adopted
8-point DHT and 8-point DCT~\cite{Loeffler1989,hou1987fastdht}.
Consequently,
further multiplicative savings are mathematically
impossible
and
because of the maturity of such methods,
even non-multiplicative savings
are very hard to be obtained.
In addition,
sinusoidal discrete transforms operate
in the real or complex number fields~\cite{Blahut2010}.
Then,
fast algorithms are often
developed for floating-point arithmetic architectures~\cite{Rizkalla2002floatingdct},
which possess relatively higher hardware cost, slower implementation when compared to fixed-point schemes~\cite{Kim1994floatingfixed}.

In the above scenario,
integer approximate transforms
have been focused to further reduce the sinusoidal transforms computational complexity overhead~\cite{britanak2007discrete, Cintra2018chapter}.
Approximate transforms
aim at preserving useful properties of the exact transformations
while favoring low-complexity computation~\cite{cintra2011integer}.
In fact,
they are not
subject to the theoretical lower bounds of
arithmetic cost
and
allow
fixed-point arithmetic
implementation~\cite{madanayake2015low,oliveira2017low}.
Furthermore,
the dyadic rational representation can be
considered to convert integer multiplication
in a combination of hardware-friendly additions and
bit-shifting operations~\cite{Lee1997csd,britanak2007discrete}.
As a consequence,
multiplierless integer approximations can be achieved
and
several methods for
the sinusoidal transforms
have been proposed~\cite{haweel2001,lengwehasatit2004scalable,bas2008,cb2011,bc2012,
	bas2013,cintra2011integer,Cintra2014-sigpro,madanayake2015low,suarez2014multi,kulasekera2015multi,Pauchard2015,da2016orthogonal, Cintra2018chapter,kouadria2013low},
being an open field of research.

Multidimensional discrete transforms
are extended versions of discrete transforms
applied to arrays with more than one dimension
and
are important tools for
multidimensional digital signal processing applications,
such as
image filtering~\cite{Gonzalez2001},
image and video coding~\cite{bhaskaran1997,Rao2001,Nguyen2011,coutinho2017low},
visual tracking~\cite{li2013visualtracking},
and
facial expression recognition~\cite{xue2015automatic}.
In particular,
multidimensional DHT-based algorithms
were successfully employed
in
medical image compression~\cite{sunder2006medical,duleba1999hartley,sunder2005performance,papitha2013compression}.

Some multidimensional discrete transforms,
such as the multidimensional DFT~($m$D~DFT)
and
the multidimensional DCT~($m$D~DCT),
satisfy the kernel separability property~\cite{Gonzalez2001},
which
allows the multidimensional computation
by taking several
one-dimensional transformation of input data~\cite{song2013local}.
Such method is called \emph{row-column} approach~\cite[p.~471]{Gonzalez2001}
and enable the fast multidimensional computation by using a fast one-dimensional algorithm.
The multidimensional DHT~($m$D~DHT) does not present such property
and,
to circumvent this characteristic,
an intermediate separable transformation
is defined,
which can be converted
into the
$m$D~DHT~\cite{bracewell1986fast}.
Then, a fast algorithm for the 1D~DHT
can be applied to compute a $m$D~DHT at the cost of additional arithmetic operations.
Hereafter,
we refer to the one-dimensional DHT simply as ``DHT''.

In recent years,
several 1D and 2D DCT approximations
were proposed
and a comprehensive review on DCT approximations
is given in \cite{Cintra2018chapter}.
In~\cite{coutinho2017low},
an approach to compute multidimensional DCT approximations based on tensor formalism~\cite{lathauwer2000best,chien2018tensor} and the separability property was proposed.
However,
to the best of our knowledge,
only in~\cite{bas2013}
and
in~\cite{cintra2011integer}
there were presented approaches
to derive DHT approximations with general block lengths
based on the Walsh-Hadamard transform and integer functions, respectively,
and approximate DHT schemes still are an unexplored field of research.
Also,
an extended approach for the approximate 3D~DHT case which deals with the non-separability property of the exact DHT lacks an algebraic formalization.
In~\cite{sunder2006medical},
Sunder~\emph{et~al.}
proposed a lossy image compression scheme
that employs the exact 3D~DHT
for \emph{Digital Imaging and Communications in Medicine}~(DICOM) data coding,
which presents better compression performance than the 3D~DFT and 3D~DCT.
However,
low-complexity computation for the 3D~DHT was not addressed.

Concurrently, emerging low-power biomedical systems, such as implantable~\cite{yakovlev2012implantable} and miniaturized biomedical devices~\cite{descour2002toward},
present severe energy and resources constraints~\cite{gao2011low}.
In the present work,
we propose a series of new 8-point DHT approximations
based on the dyadic representation.
The obtained transforms are designed to allow low-complexity multiplierless
computation.
We address two cases:
(i)~the involutional approximations approach, which applies the same algorithm for the direct and inverse
transformations; and
(ii)~the non-involutional approximations approach,
which considers a different transformation for the inverse case in order to improve
performance.
We apply the tensor formalism
to extend the 8-point
approximations to
3D~case
and several
3D~DHT approximations based on the suggested approximate transformation matrices are proposed.
We aim at employing the derived 3D~DHT approximations in
the 3D~DHT-based DICOM
lossy
data compression algorithm
proposed in~\cite{sunder2006medical}.
In addition,
we embed the proposed approximations in an ARM Cortex-M0+ core
employing the recently introduced low-cost Raspberry Pi Pico board,
which is equipped with a RP2040 microcontroller~\cite{rasp_pipico}.
We aim at analyzing the execution time and memory usage of the proposed methods.

The paper unfolds as follows.
Section~\ref{sec_1d_dht} reviews the DHT and details the proposed DHT approximations.
In Section~\ref{sec_3d_dht},
we discuss and develop the mathematical concepts to extend the proposed approximations to the three-dimensional case.
In Section~\ref{sec_dicom},
we apply the derived 3D~DHT approximations in the DICOM data compression scheme proposed in~\cite{sunder2006medical}.
In Section~\ref{sec_time_mem_complexity},
we implemented the proposed 3D~DHT approximations in
an ARM Cortex-M0+ core and the execution time and memory usage are analyzed.
Section~\ref{sec_conc} summarizes the conclusions.

\section{DHT: Mathematical Background and Proposed Approximate Methods}
\label{sec_1d_dht}

In this section,
we cover one-dimensional DHT approximations.
We review the basic mathematical concepts in Section~\ref{subsec_1d_dht}
and previous developments in the field of discrete transform approximation
in Section~\ref{subsec_1d_approx}.
Also,
we propose a set of DHT approximations in Section~\ref{subsec_1d_dht_approx}.

\subsection{One-Dimensional DHT}
\label{subsec_1d_dht}

The 1D~DHT
maps a discrete $N$-point signal
$\mathbf{x}
=
\begin{bmatrix}
x[0] & x[1] & \cdots & x[N-1]
\end{bmatrix}^\top
$
into the signal
$\mathbf{y} =
\begin{bmatrix}
y[0] & y[1] & \cdots & y[N-1]
\end{bmatrix}^\top
$
according to the following relation~\cite{bracewell1983discrete}:
\begin{align}
	\label{eq_dht_direct}
	\begin{split}
		y[k]
		\triangleq &
		\sum_{n=0}^{N-1}
		x[n]
		\cdot
		\cas
		\left(
		\frac{2\pi kn}{N}
		\right)
		,
		\quad
		k= 0,1,\ldots, N-1
		,
	\end{split}
\end{align}
where
$\cas(\cdot) \triangleq \cos (\cdot) + \sin (\cdot)$.
The inverse transformation is
equivalent to
\eqref{eq_dht_direct}, except for a scaling factor of $1/N$.
Thus,
the inverse 1D~DHT is given by
\begin{align}
\label{eq_dht_direct_2}
\begin{split}
x[n]
\triangleq &
\frac{1}{N} \cdot
\sum_{k=0}^{N-1}
y[k]
\cdot
\cas
\left(
\frac{2\pi kn}{N}
\right)
,
\quad
n= 0,1,\ldots, N-1
,
\end{split}
\end{align}
In matrix formalism,
the
DHT
can be represented by the
DHT
matrix~$\mathbf{H}_{(N)}$,
whose entries are given by:
\begin{align}
	\label{eq_dht_kernel_matrix}
	h_N[k,n]
	=
	\cas
	\left(
	\frac{2\pi kn}{N}
	\right),
	\qquad k,n = 0,1,\ldots,N-1.
\end{align}
The matrix $\mathbf{H}_{(N)}$
is orthogonal and symmetric.
Thus,
the transformation is an involution~\cite[p.~165]{Bernstein2009} and
the following property holds:
$
	\mathbf{H}_{(N)}^{-1} = \frac{1}{N} \cdot \mathbf{H}_{(N)}^\top = \frac{1}{N} \cdot \mathbf{H}_{(N)}.
$
Thus, the direct and inverse transformations are, respectively, given by
$
\mathbf{y}
=
\mathbf{H}_{(N)}
\cdot
\mathbf{x}$
and
$\mathbf{x}
=
\frac{1}{N} \cdot
\mathbf{H}_{(N)}
\cdot
\mathbf{y}$.

For $N=8$,
the DHT matrix $\mathbf{H}_{(8)}$
can
be parametrically written according to
\begin{align}
	\label{eq_dht_param}
	\mathbf{H} \left( \A, \B \right) =
	\left[ \begin{smallmatrix}
		\A &  \A &  \A &  \A      &  \A &  \A &  \A &  \A \\
		\A &  \B &  \A &   0      & -\A & -\B & -\A &   0 \\
		\A &  \A & -\A & -\A      &  \A &  \A & -\A & -\A \\
		\A &   0 & -\A &  \B      & -\A &   0 &  \A & -\B \\
		\A & -\A &  \A & -\A      &  \A & -\A &  \A & -\A \\
		\A & -\B &  \A &   0      & -\A &  \B & -\A &   0 \\
		\A & -\A & -\A &  \A      &  \A & -\A & -\A &  \A \\
		\A &   0 & -\A & -\B      & -\A &   0 &  \A &  \B
	\end{smallmatrix} \right],
\end{align}
where
$\A = 1$ and
$\B = \sqrt{2}$.
Such matrix can be factorized according to:
\begin{align}
	\mathbf{H}\left( \A, \B \right)
	=
	\mathbf{A}_3
	\cdot
	\mathbf{A}_2
	\cdot
	\mathbf{M}\left( \A, \B \right)
	\cdot
	\mathbf{A}_1
	\cdot
	\mathbf{P},
	\label{eq_dht_factorization}
\end{align}
where
$\mathbf{P}$ is a permutation matrix
and $\mathbf{M}\left(\A,\B \right)$ is a matrix with multiplicand elements, respectively
given by
\begin{align*}
	\mathbf{P}
	=
	\left[ \begin{smallmatrix}
		1 & 0 & 0 & 0    & 0 & 0 & 0 & 0 \\
		0 & 0 & 0 & 0    & 1 & 0 & 0 & 0 \\
		0 & 0 & 1 & 0    & 0 & 0 & 0 & 0 \\
		0 & 0 & 0 & 0    & 0 & 0 & 1 & 0 \\
		0 & 1 & 0 & 0    & 0 & 0 & 0 & 0 \\
		0 & 0 & 0 & 0    & 0 & 1 & 0 & 0 \\
		0 & 0 & 0 & 1    & 0 & 0 & 0 & 0 \\
		0 & 0 & 0 & 0    & 0 & 0 & 0 & 1 \\
	\end{smallmatrix} \right]
	\;\text{ and }\;
	\mathbf{M}\left(\A,\B \right)
	=
	\left[ \begin{smallmatrix}
		\A &   &   &      &   &   &   &   \\
		& \A &   &      &   &   &   &   \\
		&   & \A &      &   &   &   &   \\
		&   &   & \A    &   &   &   &   \\
		&   &   &      & \A &   &   &   \\
		&   &   &      &   & \B&   &   \\
		&   &   &      &   &   & \A &   \\
		&   &   &      &   &   &   & \B\\
	\end{smallmatrix} \right],
\end{align*}
$\mathbf{A}_1$,
$\mathbf{A}_2$,
and
$\mathbf{A}_3$
are additive sparse matrices,
respectively given by
\begin{align*}
	\mathbf{A}_1
	=
	\left[ \begin{smallmatrix}
		\mathbf{B} & & & \\
		&\mathbf{B} & & \\
		&  &\mathbf{B} & \\
		& &  & \mathbf{B}
	\end{smallmatrix} \right],
	\quad
	\mathbf{A}_2
	=
	\left[ \begin{smallmatrix}
		1 & \phantom{-}0 & \phantom{-}1 & \phantom{-}0 & & & & \\
		0 & \phantom{-}1 & \phantom{-}0 & \phantom{-}1 & & & & \\
		1 & \phantom{-}0 & -1 & \phantom{-}0 & & & & \\
		0 & \phantom{-}1 & \phantom{-}0 & -1 & & & & \\
		& & & & 1 & \phantom{-}0 & \phantom{-}1 & \phantom{-}0  \\
		& & & &  0 & \phantom{-}1 & \phantom{-}0 & \phantom{-}0 \\
		& & & & 1 & \phantom{-}0 & -1 & \phantom{-}0  \\
		& & & & 0 & \phantom{-}0 & \phantom{-}0 & \phantom{-}1  \\
	\end{smallmatrix} \right],
	\quad
	\text{and}
	\quad
	\mathbf{A}_3
	= &
	\begin{bmatrix}
		\mathbf{I}_4 & \phantom{-}\mathbf{I}_4 \\
		\mathbf{I}_4 & -\mathbf{I}_4
	\end{bmatrix},
\end{align*}
$\mathbf{B}= \left[ \begin{smallmatrix}
1 &  \phantom{-}1 \\
1 & -1
\end{smallmatrix} \right]$,
$\mathbf{I}_4$ is the identity matrix of size $4\times 4$,
and the omitted entries are zero.
Such factorization
is obtained by
employing the Sorensen split-radix algorithm~\cite{sorensen1985computing}
with $N=8$.
The factorization can be verified by multiplying the matrices
in~\eqref{eq_dht_factorization} and comparing the result with \eqref{eq_dht_param}.

To calculate the arithmetic complexity,
we need to count the number of additions and multiplications
required to perform the whole matrix product in \eqref{eq_dht_factorization}.
Notice that any matrix row containing $r$ ones and $8-r$ zeros
imposes $r-1$ additions to the arithmetic complexity.
Also, a diagonal matrix does not introduce any addition,
only multiplications by the diagonal elements.
Finally, the permutation matrix does not
require any arithmetic operation,
since it is just data reorganization
--- it can be implemented in hardware by appropriate wiring and in software by permuting code lines.
Thus:
\begin{itemize}
\item the matrix $\mathbf{P}$ does not introduce
any arithmetic operation;
\item the matrix $\mathbf{A}_1$ presents six rows with two '1's and two rows with a single '1',
introducing only 6 additions;
\item the matrix $\mathbf{A}_2$ presents eight rows with two '1's,
introducing 8 additions;
\item the matrix $\mathbf{A}_3$ presents eight rows with two '1's,
introducing 8 additions;
\item the matrix $\mathbf{M}\left( \A, \B \right)$ presents six diagonal elements $\A$ and two diagonal elements $\B$.
However, since $\A=1$, only 2 multiplications by $\B = \sqrt{2}$ are required.
\end{itemize}
Consequently,
such factorization results in an arithmetic complexity of
22 additions and 2 multiplications.

\subsection{Sinusoidal Transform Approximations}
\label{subsec_1d_approx}

An approximation of a discrete transformation
aims at behaving similarly to the exact transformation
according to some criterion~\cite{Cintra2018chapter}.
An approximate transform of size~$N$
is represented
by an $N \times N$ matrix
$\hat{\mathbf{T}}_{(N)}$,
which preserves
useful
properties of
a given exact
transformation matrix
$\mathbf{T}_{(N)}$.
Exact and approximate transforms are usually compared by
quantitative metrics,
such as
the mean squared error~(MSE)~\cite[p.~162]{britanak2007discrete}
and
the coding gain~$C_g$~\cite[p.~163]{britanak2007discrete}.
Furthermore,
an approximation is typically designed to present low-complexity computation.
Generally,
exact sinusoidal transforms present
transformation matrices with real-or complex-valued entries
and the
approximate matrices are derived
with
selected numerical values,
such as the dyadic rationals,
which are suitable for
hardware implementation~\cite{gerek20062}.
A dyadic rational is a number represented in
the form $l/2^{p}$,
where $l$
and $p$ are integers and $l$ is odd~\cite[p.~181]{britanak2007discrete}.
By considering the canonic signed representation~(CSD)~\cite{Lee1997csd},
the product of an arbitrary integer by a dyadic rational
can be converted into additions and bit-shifting operations,
which present noticeable faster hardware implementations than
usual integer multiplication algorithms~\cite{bas2008, oliveira2017jpeg}.

The approximate 1D~transformation of a vector $\mathbf{x}$ is the transform-domain vector $\mathbf{y}$, computed according to:
\begin{align}
	\mathbf{y} &= \hat{\mathbf{T}}_{(N)} \cdot \mathbf{x}.
	\label{eq_1d_sinu_approx}
\end{align}
If the approximate matrix is orthogonal,
then the product
$\hat{\mathbf{T}}_{(N)} \cdot {\hat{\mathbf{T}}_{(N)}}^\top$
results in a diagonal matrix~\cite{Cintra2014-sigpro}.
Let
\begin{align}
	\mathbf{D}_{(N)} = \left( \hat{\mathbf{T}}_{(N)} \cdot {\hat{\mathbf{T}}_{(N)}}^\top \right)^{-1},
	\label{eq_matrix_D}
\end{align}
which is also a diagonal matrix.
Then,
$\hat{\mathbf{T}}_{(N)}^{-1} =
{\hat{\mathbf{T}}_{(N)}}^\top
\cdot
\mathbf{D}_{(N)}
$
and
the inverse transformation of \eqref{eq_1d_sinu_approx}
is given by
$
	\mathbf{x} =
	{\hat{\mathbf{T}}_{(N)}}^\top
	\cdot
	\mathbf{D}_{(N)}
	\cdot
	\mathbf{y}
$.

Non-orthogonal approximations are also found in literature~\cite{haweel2001,Cintra2014-sigpro,cintra2011integer}.
In this context,
approximate transforms
can present the quasi-orthogonality property,
in which
$\hat{\mathbf{T}}_{(N)} \cdot {\hat{\mathbf{T}}_{(N)}}^\top$
is ``almost'' a diagonal matrix.
The \emph{deviation from diagonality}
metric~$\delta(\cdot )$
was proposed in~\cite{Cintra2014-sigpro},
allowing the quantification of quasi-orthogonality.
This concept can be ,
as shown in section 2.1.
transported to
quasi-orthogonal
approximations:
$\mathbf{D}_{(N)} = \operatorname{diag}\left( \left( \hat{\mathbf{T}}_{(N)} \cdot {\hat{\mathbf{T}}_{(N)}}^\top \right)^{-1} \right)$~\cite{Cintra2014-sigpro},
where $\operatorname{diag}(\cdot)$
returns a diagonal matrix
with diagonal elements of its matrix argument.
In the context of image compression,
diagonal matrices do not introduce any
computational overhead since they can be merged in the quantization or dequantization stages of image coding schemes~\cite{britanak2007discrete}.

\subsection{Proposed 8-point DHT Approximations}
\label{subsec_1d_dht_approx}

In this section,
we focus on approximating the 8-point DHT matrix $\mathbf{H}$.
First,
we notice that
if
an approximation preserves
the parametric matrix structure
of the exact transformation~\eqref{eq_dht_param},
then
it
also shares
the same factorization in~\eqref{eq_dht_factorization}.
Thus,
the approximate matrix
can be written according to
$
	\hat{\mathbf{H}} \triangleq
	\mathbf{H} \left(\hat{\A}, \hat{\B} \right)
$.
It is possible to show that the orthogonality
is satisfied
only if
$
	\hat{\B} = \sqrt{2} \cdot \hat{\A}
$.
Consequently,
it is not possible to find
an orthogonal approximation
$\hat{\mathbf{H}}$
with
rational elements for
both
$\hat{\A}$ and $\hat{\B}$
preserving such parametric structure.
Then,
we direct our attention to
non-orthogonal transformations presenting
quasi-orthogonality properties~\cite{Cintra2014-sigpro}.

Our goal is to obtain multiplierless DHT approximations with
low computational cost,
high coding gain,
low MSE,
and quasi-orthogonality.
Furthermore,
we also focus on
obtaining low-complexity inverse transformations,
which can benefit applications with power and speed constraints for both the encoder and the decoder sides.
Then,
we separate the problem in two cases:
(i)~involutional approach,
in which
we consider the direct and inverse transformations based on the same matrix,
maintaining the exact DHT involution property,
which benefits applications that need to use the same hardware to encode and decode data,
and exploiting the low-complexity direct algorithm for the inverse computation;
and
(ii)~non-involutional approach,
in which we relax the involution property,
allowing the direct
and inverse matrices to be different in order to find a low-complexity matrix that better approximates the inverse transformation according to
the deviation from diagonality.

\begin{figure}
	\centering
	\includegraphics[scale=1]{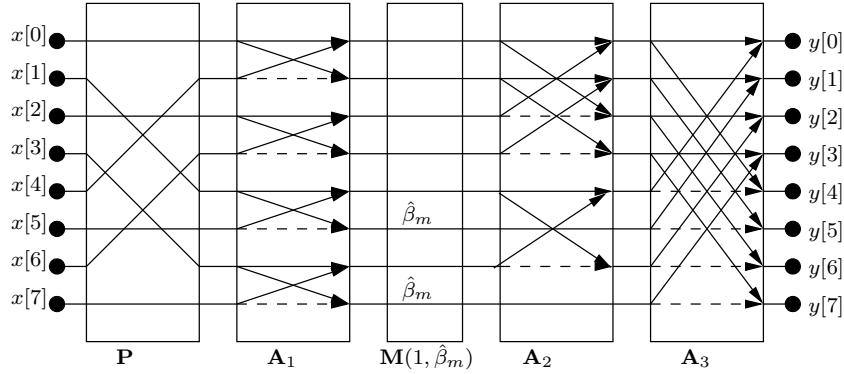}
	\caption{
		Fast algorithm for the 8-point DHT. Dashed lines represent multiplications by -1.
		}
	\label{fig_fast_alg}
\end{figure}

We preserve the parameter $\hat{\A }=\A = 1$
because it is a trivial multiplicand element.
An exhaustive search is made for the parameter $\hat{\B}$,
ranging $\hat{\B} \in \left[0, 3\right]$ at dyadic steps of $1/8$,
where the exact value $\B = \sqrt{2}$ is roughly in the middle of such interval.
Such dyadic step is chosen to ensure
that the resulting transformations
possess low additive complexities.
Indeed,
it is possible to show that such dyadic step and interval guarantee dyadic rationals with no more than 3 extra additions.
Then, $\hat{\B}_m = \frac{m}{8}$ for $m=1,2,\ldots,24$.
We exclude $\hat{\B}_0$ because such value generates a
singular matrix.
With $\hat{\A}=1$ and $\hat{\B}_m$,
it is generated a new matrix
parametrically written as a function of $\hat{\B}_m$ according to:
\begin{align}
	\hat{\mathbf{H}} \left( {\hat{\B}_m} \right)
	=
	\mathbf{A}_3
	\cdot
	\mathbf{A}_2
	\cdot
	\mathbf{M}\left( 1, \hat{\B}_m \right)
	\cdot
	\mathbf{A}_1
	\cdot
	\mathbf{P},
	\label{eq_dht_factorization_approx}
\end{align}
which implies an approximate transform given by
\begin{align}
\begin{split}
\mathbf{y}  & =  \left[ \hat{\mathbf{H}} \left( {\hat{\B}_m} \right) \right] \cdot \mathbf{x}
\\
 &=
   \left[ \mathbf{A}_3
 \cdot
 \mathbf{A}_2
 \cdot
 \mathbf{M}\left( 1, \hat{\B}_m \right)
 \cdot
 \mathbf{A}_1
 \cdot
 \mathbf{P} \right] \cdot \mathbf{x}.
\end{split}
\label{eq_dht_factorization_approx_full}
\end{align}
The fast algorithm for such factorization is shown in \figurename~\ref{fig_fast_alg},
where the input and output coefficients are the entries of vectors $\mathbf{x}$ and $\mathbf{y}$
in~\eqref{eq_dht_factorization_approx_full},
respectively.
Each stage in the fast algorithm in Figure~\ref{fig_fast_alg} corresponds to
a matrix product in the factorization given in~\eqref{eq_dht_factorization_approx_full}.
Two arrows joining the same node means an addition operation,
dashed lines correspond to multiplications by -1,
and the multiplication by the coefficients $\hat{\B}_m$
are explicit.
The total additive complexity depends on the additive complexity associated with the multiplication by the term $\hat{\B}_m$ in the dyadic representation.
It requires 22 additions plus twice the complexity associated with $\hat{\B}_m$.

First considering the (i)~involutional approach,
in which we employ the same matrix for direct and inverse transformation,
we optimized the approximation search according to the
deviation from diagonality
$\delta \left[ \hat{\mathbf{H}}\left(\hat{\B}_m \right) \cdot \hat{\mathbf{H}}\left(\hat{\B}_m \right) \right]$,
the MSE,
the coding gain,
and the computational complexity.
Then,
for the (ii)~non-involutional approach,
we employed the same parametric search for
both direct and inverse transformations.
In this case,
we allow the direct and inverse transformations
to be different.
For each obtained
approximate DHT matrix
$\hat{\mathbf{H}}\left(\hat{\B}_m \right)$,
we search for an approximate inverse matrix
$\hat{\mathbf{H}}\left(\hat{\B}_q \right)$
which minimizes the deviation from diagonality
$\delta \left[ \hat{\mathbf{H}}\left(\hat{\B}_m \right) \cdot \hat{\mathbf{H}}\left(\hat{\B}_q \right) \right]$.

The matrix~$\hat{\mathbf{H}}\left(\hat{\B}_{11} \right)$ presents
the lowest MSE value;
the matrix~$\hat{\mathbf{H}}\left(\hat{\B}_{12} \right)$ shows
the highest coding gain;
and
the matrix~$\hat{\mathbf{H}}\left(\hat{\B}_8 \right)$
possesses the lowest computational cost.
For the non-involutional approach,
we found that
the deviation from diagonality decreases when
the matrix~$\hat{\mathbf{H}}\left(\hat{\B}_{12}\right)$
is taken
as
the
quasi-inverse for $\hat{\mathbf{H}}\left(\hat{\B}_{11} \right)$ and vice-versa.
Also,
the matrix $\hat{\mathbf{H}}\left(\hat{\B}_{8} \right)$
presents the matrix $\hat{\mathbf{H}}\left(\hat{\B}_{16} \right)$
as its exact inverse,
i.e.,
$\delta \left[ \hat{\mathbf{H}}\left(\hat{\B}_{8} \right) \cdot \hat{\mathbf{H}}\left(\hat{\B}_{16} \right) \right] = 0$.
Table~\ref{tab_beta_parameters}
shows the numerical representation and complexity
for each parameter~$\B_m$
of the derived matrices.
Figure-of-merit measurements are summarized in Table~\ref{tab_metrics}.

\begin{table}
	\centering
	\caption{Representation and complexity of multiplying $\B_m$ parameters of each derived DHT approximate matrix (only additions and shifts are required)}
	\label{tab_beta_parameters}
	\begin{tabular}{l c  c  c c}
		\toprule
		\multirow{2}{*}{$m$}& \multirow{2}{*}{$\B_m$} & \multirow{2}{*}{CSD representation} & \multicolumn{2}{c}{Complexity} \\
		\cmidrule(r){4-5}
		& &  & Addition & Shift \\
		\cmidrule(r){1-5}
		$8$ & $1$ & $1$ & 0 & 0  \\
		$11$ & ${11}/{8}$ & $1 + \frac{1}{4} + \frac{1}{8}$ & 2 & 2  \\
		$12$ & ${3}/{2}$ & $1 + \frac{1}{2}$ & 1 & 1  \\
		$16$ & $2$ & $2$ & 0 & 1  \\
		\bottomrule
	\end{tabular}
\end{table}

\begin{table*}
	\centering
	\caption{Invertible and quasi-inverses DHT matrices Approximations}
	\label{tab_metrics}
	\begin{tabular}{c  c | c  c  c | c c }
		\toprule
		\multicolumn{1}{c}{Direct} &
		\multicolumn{1}{c}{Inverse} &
		\multicolumn{3}{c}{Metrics} &
		\multicolumn{2}{c}{$\hat{\mathbf{H}}\left(\hat{\B}_m \right)$~complexity}
		\\
		\cmidrule(r){1-7}
		$\hat{\mathbf{H}}\left(\hat{\B}_m \right)$ &
		$\hat{\mathbf{H}}\left(\hat{\B}_q \right)$ &
		$\delta \left[ \hat{\mathbf{H}}\left( \hat{\B}_m \right) \cdot \hat{\mathbf{H}}\left(\hat{\B}_q \right) \right]$ & $C_g$~(dB) & MSE&
		Addition & Shift
		\\
		\midrule
		$\hat{\mathbf{H}}\left(1 \right)$ &
		$\hat{\mathbf{H}}\left(1 \right)$ &
		$1.94\cdot 10^{-2}$  & $ 7.418$  & $3.182 \cdot 10^{-2}$
		& $\mathbf{22}$ & $\mathbf{0}$ \\
		$\hat{\mathbf{H}}\left(\frac{11}{8} \right)$ &
		$\hat{\mathbf{H}}\left(\frac{11}{8} \right)$ &
		${1.92\cdot 10^{-4}}$  & $7.818$  & $\mathbf{2.85\cdot 10^{-4}}$
		& 24 & 2\\
		$\hat{\mathbf{H}}\left(\frac{3}{2} \right)$ &
		$\hat{\mathbf{H}}\left(\frac{3}{2} \right)$ &
		$9.16\cdot 10^{-4}$  & $\mathbf{7.830}$  &  $13.65\cdot 10^{-4}$
		& 23 & 1 \\

		$\hat{\mathbf{H}}\left(1 \right)$ &
		$\hat{\mathbf{H}}\left(2 \right)$ &
		$\mathbf{0}$ & $ 7.418$  & $3.182 \cdot 10^{-2}$
		& $\mathbf{22}$ & $\mathbf{0}$ \\
		$\hat{\mathbf{H}}\left(\frac{11}{8} \right)$ &
		$\hat{\mathbf{H}}\left(\frac{3}{2} \right)$ &
		$6.01\cdot 10^{-5}$  & $7.818$  & $\mathbf{2.852\cdot 10^{-4}}$
		& 24 & 2\\
		$\hat{\mathbf{H}}\left(\frac{3}{2} \right)$ &
		$\hat{\mathbf{H}}\left(\frac{11}{8} \right)$ &
		$6.01\cdot 10^{-5}$  & $\mathbf{7.830}$  &  $1.365\cdot 10^{-3}$
		& 23 & 1\\
		$\hat{\mathbf{H}}\left(2 \right)$ &
		$\hat{\mathbf{H}}\left(1 \right)$ &
		$\mathbf{0}$ & $7.506$  & $6.365 \cdot 10^{-2}$
		& 22 & 1\\
		\bottomrule
	\end{tabular}
\end{table*}

\section{Three-Dimensional DHT:
	High-order Tensor,
	Mathematical Definition
	and Proposed Approximate Formalism}
\label{sec_3d_dht}

In the current Section,
we cover the exact and approximate 3D~DHT mathematical definition.
We develop the exact 3D~DHT definition based on
high-order tensor formalism in Section~\ref{subsec_3d_dht}.
In Section~\ref{subsec_3d_dht_approx},
we
propose a general algebraic definition for the approximate 3D~case based on DHT approximate matrices,
where
we applied the proposed 8-point DHT approximations derived in the previous Section~\ref{subsec_1d_dht_approx}
to generate new $8\times 8 \times 8$ 3D~DHT approximations.
We also discuss their arithmetic complexity in Section~\ref{subsec_complex_assess}.

\subsection{High-order Tensor and Exact Three-dimensional DHT}
\label{subsec_3d_dht}

In a general way,
an $R$th-order tensor can be understood as an
array with $R$ indices~\cite{lathauwer1998}.
In particular,
a vector and a matrix are first- and second-order tensor, respectively.
Let
$\mathcal{A}$ be an
$N_1 \times N_2 \times \cdots \times N_R$ $R$th-order tensor,
whose entries are given by
$a[n_1,n_2,\ldots,n_R]$ and
$n_i = 0,1,\ldots,N_i-1,$ for $i=1,2,\ldots, R$.
The $i$-mode product of such tensor by
an $H \times N_i$ matrix $\mathbf{M}$,
denoted by $\mathcal{A} \times_i \mathbf{M}$,
is a tensor
$\mathcal{B}$
of size
$N_1 \times \cdots \times
N_{i-1} \times H \times N_{i+1} \times \cdots \times N_R$,
whose entries are defined as~\cite{lathauwer2000best}:
\begin{align}
	\label{eq_imode}
	\begin{split}
		b[n_1, \ldots, n_{i-1}, h, n_{i+1}, \ldots, n_R]
		\triangleq &
		\sum_{n_i=0}^{N_i-1}
		a[n_1, \ldots, n_i, \ldots, n_R]
		\cdot
		m[h,n_i]
		,
	\end{split}
\end{align}
where $m[h,n_i]$ are the matrix $\mathbf{M}$ entries and $h=0,1,\ldots,H-1$.

Taking into account a matrix
$\mathbf{H}$ of size $N_i \times L$,
a matrix
$\mathbf{D}$ of size $L \times N_i$,
and
an $R$th-order tensor
$\mathcal{T}$ of size $N_1 \times N_2 \times \cdots \times N_i \times \cdots \times N_R$,
it can be shown that~\cite{lathauwer2000best}:
\begin{align}
	\mathcal{T} \times_i \left( \mathbf{H} \cdot \mathbf{D} \right)
	=
	\mathcal{T} \times_i \mathbf{D} \times_i \mathbf{H}.
	\label{eq_porp_dist}
\end{align}

The 3D~DHT of a third-order tensor $\mathcal{X}$ of
size $N_1 \times N_2 \times N_3$,
whose entries are $x[n_1,n_2,n_3]$,
for $n_i = 0,1,\ldots, N_i-1$ and $i=1,2,3$,
is
defined as
the transform-domain third-order tensor $\mathcal{Y}$,
whose entries are given by:
\begin{align}
	\begin{split}
		y[k_1,k_2,k_3]
		\triangleq &
		\sum_{n_1=0}^{N_1-1}
		\sum_{n_2=0}^{N_2-1}
		\sum_{n_3=0}^{N_3-1}
		x[n_1,n_2,n_3]
		\cdot
		\cas
		\left(
		\frac{2\pi n_1 k_1}{N_1} +
		\frac{2\pi n_2 k_2}{N_2} +
		\frac{2\pi n_3 k_3}{N_3}
		\right),
		\\
		&
		k_i = 0,1,\ldots, N_i-1,
		\qquad i=1,2,3.
	\end{split}
	\label{eq_3d_dht_sum}
\end{align}
Similarly to the 1D case,
the inverse transformation is identical
to the direct one,
except for a scaling factor of
${1}/{\left( N_1 N_2 N_3 \right)}$,
i.e.,
it is given by
\begin{align}
\begin{split}
x[n_1,n_2,n_3]
\triangleq &
\frac{1}{N_1 N_2 N_3}
\sum_{k_1=0}^{N_1-1}
\sum_{k_2=0}^{N_2-1}
\sum_{k_3=0}^{N_3-1}
y[k_1,k_2,k_3]
\cdot
\cas
\left(
\frac{2\pi n_1 k_1}{N_1} +
\frac{2\pi n_2 k_2}{N_2} +
\frac{2\pi n_3 k_3}{N_3}
\right),
\\
&
n_i = 0,1,\ldots, N_i-1,
\qquad i=1,2,3.
\end{split}
\label{eq_3d_dht_sum_2}
\end{align}

In contrast to
the 3D~DCT and the 3D~DFT,
the 3D~DHT does not present the separability property~\cite{Gonzalez2001}.
To overcome such behavior of the 3D~DHT,
it is defined the \emph{special}
3D~DHT~(3D~SDHT)~\cite{watson1986separable,mandal2013separable},
which is separable and given by a tensor $\mathcal{Y}_s$ of size $N_1 \times N_2 \times N_3$,
whose entries are defined as
\begin{align}
	\begin{split}
		y_s[k_1,k_2,k_3]
		\triangleq &
		\sum_{n_1=0}^{N_1-1}
		\sum_{n_2=0}^{N_2-1}
		\sum_{n_3=0}^{N_3-1}
		x[n_1,n_2,n_3]
		\cdot
		\cas
		\left(
		\frac{2\pi n_1 k_1}{N_1}
		\right)
		\cdot
		\cas
		\left(
		\frac{2\pi n_2 k_2}{N_2}
		\right)
		\cdot
		\cas
		\left(
		\frac{2\pi n_3 k_3}{N_3}
		\right)
		,
		\\
		=&
		\sum_{n_1=0}^{N_1-1}
		x[n_1,n_2,n_3]
		\cdot
		\cas
		\left(
		\frac{2\pi n_1 k_1}{N_1}
		\right)
		\cdot
		\sum_{n_2=0}^{N_2-1}
		\cas
		\left(
		\frac{2\pi n_2 k_2}{N_2}
		\right)
		\cdot
		\sum_{n_3=0}^{N_3-1}
		\cas
		\left(
		\frac{2\pi n_3 k_3}{N_3}
		\right),
		\qquad
		\\ &
		k_i = 0,1,\ldots,N_i-1,
		\qquad
		i=1,2,3.
	\end{split}
	\label{eq_3d_dht_spec1}
\end{align}
It is possible to show that
the 3D~SDHT can be expressed in terms of
$i$-mode products~\eqref{eq_imode}
by the DHT matrix in each dimension~\eqref{eq_dht_kernel_matrix}
according to:
\begin{align}
	\mathcal{Y}_s = \mathcal{X} \times_1 \mathbf{H}_{(N_1)} \times_2 \mathbf{H}_{(N_2)} \times_3 \mathbf{H}_{(N_3)}.
	\label{eq_3d_dht_spec1_tensor}
\end{align}
Considering that~\cite{hao1987three}
\begin{align*}
	\begin{split}
		\cas(\alpha + \beta + \gamma)
		= &
		\frac{1}{2}
		[
		\cas (-\alpha) \cdot \cas (\beta) \cdot \cas (\gamma)
		+
		\cas (\alpha) \cdot \cas (-\beta) \cdot \cas (\gamma)
		\\ &
		+
		\cas (\alpha) \cdot \cas (\beta)
		\cdot \cas (-\gamma)
		-
		\cas (-\alpha) \cdot \cas (-\beta) \cdot \cas (-\gamma)
		],
	\end{split}
\end{align*}
the 3D~DHT tensor
$\mathcal{Y}$
can be expressed
in terms of
rearranged versions
of
3D~SDHT
tensor $\mathcal{Y}_s$
according to
$
	\mathcal{Y} =
	\frac{1}{2}
	\left(
	\mathcal{Y}_s^{(1)}
	+
	\mathcal{Y}_s^{(2)}
	+
	\mathcal{Y}_s^{(3)}
	-
	\mathcal{Y}_s^{(4)}
	\right)
$,
where
the entries of $\mathcal{Y}_s^{(1)}, \mathcal{Y}_s^{(2)}, \mathcal{Y}_s^{(3)}$, and $\mathcal{Y}_s^{(4)}$
are given respectively by:
\begin{align}
	\label{eq_dht_manipulation}
	\begin{split}
		y_s^{(1)}[k_1,k_2,k_3] =& y_s\left[\left((N_1-k_1)\right)_{N_1}, k_2, k_3\right], \\
		y_s^{(2)}[k_1,k_2,k_3] =& y_s\left[k_1, \left((N_2-k_2)\right)_{N_2}, k_3\right], \\
		y_s^{(3)}[k_1,k_2,k_3] =& y_s\left[k_1, k_2, \left((N_3-k_3)\right)_{N_3} \right],\\
		y_s^{(4)}[k_1,k_2,k_3] =& y_s\left[\left((N_1-k_1)\right)_{N_1}, \left((N_2-k_2)\right)_{N_2},
		\left((N_3-k_3)\right)_{N_3}\right],
		\\ &
		k_i = 0,1,\ldots,N_i-1, \qquad i=1,2,3,
	\end{split}
\end{align}
and
$((k))_N$ indicates
$\left(k \operatorname{modulo} N\right)$~\cite[p.~142]{oppenheim2010discrete}.
The inverse transformation is computed by similar procedure including the scaling factor of
$1/\left( N_1 N_2 N_3 \right)$:
(i)~first, compute the inverse 3D~SDHT, given in terms of $i$-mode products~\cite{lathauwer2000best} by:
\begin{align}
\mathcal{X}_s = \frac{1}{N_1 N_2 N_3} \cdot \mathcal{Y} \times_1 \mathbf{H}_{(N_1)} \times_2 \mathbf{H}_{(N_2)} \times_3 \mathbf{H}_{(N_3)};
\label{eq_3d_dht_spec1_tensor_2}
\end{align}
(ii)~then, compute $\mathcal{X} =
\frac{1}{2}
\left(
\mathcal{X}_s^{(1)}
+
\mathcal{X}_s^{(2)}
+
\mathcal{X}_s^{(3)}
-
\mathcal{X}_s^{(4)}
\right)
$,
where
the entries of $\mathcal{X}_s^{(1)}, \mathcal{X}_s^{(2)}, \mathcal{X}_s^{(3)}$, and $\mathcal{X}_s^{(4)}$
are given analogously by~\eqref{eq_dht_manipulation}.

\subsection{General Mathematical Formulation for the Approximate 3D~Case}
\label{subsec_3d_dht_approx}

Our objective in this section is
the derivation of a general approach
for
approximating the exact 3D~DHT,
defined in~\eqref{eq_3d_dht_sum}.
We aim at proposing a method based on
the 1D~approximate matrix formalism describe
in Section~\ref{subsec_1d_approx}
and then extend to the 3D~case.
The motivations for such approach are:
\begin{enumerate}[i)]
	\item the 3D DHT approximation would
	have a straightforward mathematical relationship
	with
	a 1D~DHT approximation version;
	\item a multidimensional fast algorithm
	based on the row-column approach
	can be proposed considering an 1D~fast algorithm;
	and
	\item one-dimensional matrix approximation techniques as the one described in Section~\ref{subsec_1d_approx} and approximate matrices as derived in Section~\ref{subsec_1d_dht_approx} can be employed to obtain 3D~methods.
\end{enumerate}

Then,
instead of approximating directly~\eqref{eq_3d_dht_sum},
we primarily approximate the SDHT in~\eqref{eq_3d_dht_spec1}
according to the tensor formalism in~\eqref{eq_3d_dht_spec1_tensor}
and~\eqref{eq_dht_manipulation}.
Let~$\mathcal{X}$ be an arbitrary
third-order tensor  of
size~$N_1 \times N_2 \times N_3$
and
$\hat{\mathbf{H}}_{(N_1)}$,
$\hat{\mathbf{H}}_{(N_2)}$,
$\hat{\mathbf{H}}_{(N_3)}$ be
arbitrary approximate DHT matrices.
We define
the 3D~SDHT by replacing the exact matrices
in~\eqref{eq_3d_dht_spec1_tensor}
for the respective approximate matrices
in each dimension, according to:
\begin{align}
	\mathcal{Y}_s & \triangleq \mathcal{X} \times_1 \hat{\mathbf{H}}_{(N_1)} \times_2 \hat{\mathbf{H}}_{(N_2)} \times_3 \hat{\mathbf{H}}_{(N_3)},
	\label{eq_3d_direct_approx}
\end{align}
where
$\mathcal{Y}_s$
is the approximate 3D~SDHT transform-domain third-order tensor.
Then, we derive the approximate 3D~DHT from the approximate 3D~SDHT
analogously to
the exact case,
according to the following:
\begin{align}
	\mathcal{Y} \triangleq
	\frac{1}{2} \cdot
	\left(
	\mathcal{Y}_s^{(1)}
	+
	\mathcal{Y}_s^{(2)}
	+
	\mathcal{Y}_s^{(3)}
	-
	\mathcal{Y}_s^{(4)}
	\right).
	\label{eq_dht_from_sdht2}
\end{align}
The inverse procedure is computed in a similar way.
First,
we
compute
\begin{align}
	\begin{split}
		\mathcal{X}_s =& \mathcal{Y} \times_1 \hat{\mathbf{H}}^{-1}_{(N_1)} \times_2 \hat{\mathbf{H}}^{-1}_{(N_2)} \times_3 \hat{\mathbf{H}}^{-1}_{(N_3)} \\
		= & \mathcal{Y}
		\times_1 \left(\hat{\mathbf{H}}_{(N_1)} \cdot \mathbf{D}_{(N_1)}  \right)
		\times_2 \left(\hat{\mathbf{H}}_{(N_2)} \cdot \mathbf{D}_{(N_2)} \right)
		\times_3 \left(\hat{\mathbf{H}}_{(N_3)} \cdot \mathbf{D}_{(N_3)} \right) \\
		=& \mathcal{Y}
		\times_1 \mathbf{D}_{(N_1)}
		\times_2 \mathbf{D}_{(N_2)}
		\times_3 \mathbf{D}_{(N_3)}
		\times_1 \hat{\mathbf{H}}_{(N_1)}
		\times_2 \hat{\mathbf{H}}_{(N_2)}
		\times_3 \hat{\mathbf{H}}_{(N_3)}
		,
	\end{split}
	\label{eq_3d_inverse_approx}
\end{align}
where $\mathbf{D}_N$ is given by~\eqref{eq_matrix_D}.
In~\eqref{eq_3d_inverse_approx},
we applied the property described in~\eqref{eq_porp_dist}.
The diagonal matrix product
can be merged into the 3D~dequantization step
and
completely eliminated
from
the transform computation step~\cite{britanak2007discrete,sunder2006medical,coutinho2017low}.
Finally,
we obtain
$
	\mathcal{X} =
	\frac{1}{2} \cdot
	\left(
	\mathcal{X}_s^{(1)}
	+
	\mathcal{X}_s^{(2)}
	+
	\mathcal{X}_s^{(3)}
	-
	\mathcal{X}_s^{(4)}
	\right)
$.

For the particular case $N_1 = N_2 = N_3 \triangleq N$,
a single approximate matrix $\hat{\mathbf{H}}_{(N)}$
can be employed to compute the 3D~DHT approximation
by considering $\hat{\mathbf{H}}_{(N_1)} = \hat{\mathbf{H}}_{(N_2)} = \hat{\mathbf{H}}_{(N_3)} \triangleq \hat{\mathbf{H}}_{(N)}$.
Thus, we have that:
\begin{align}
	\mathcal{Y}_s &= \mathcal{X} \times_1 \hat{\mathbf{H}}_{(N)} \times_2 \hat{\mathbf{H}}_{(N)} \times_3 \hat{\mathbf{H}}_{(N)}.
	\label{eq_3d_direct_approx_sym}
\end{align}
In this work,
we focus on such particular case with $N=8$.
We consider
the matrices derived in
Section~\ref{subsec_1d_dht_approx}
applied to~\eqref{eq_3d_direct_approx_sym}
to obtain
a set of 3D~DHT approximations.

\subsection{Complexity Assessment}
\label{subsec_complex_assess}

Let ${C}_{1D}\left( \mathbf{H}_N \right)$
be the arithmetic complexity of transformation
represented by $\mathbf{H}_N$ for the 1D~case~\eqref{eq_1d_sinu_approx}.
The aforesaid measure generically encompasses
the multiplicative, additive and bit-shift complexities
and it depends on the particular chosen algorithm for 1D~transformation.
In~\eqref{eq_3d_direct_approx_sym},
there are two free indices ranging in $[0,N-1]$
for each $i$-mode product~\eqref{eq_imode}.
Consequently,
there are $N^2$
1D~DHT computation~\eqref{eq_3d_direct_approx}
for each $i$-mode product.
Then,
the 3D~SDHT arithmetic complexity ${C}_{3D}\left( \mathbf{H}_N \right)$ can be obtained from the 1D~case
according to:
\begin{align}
	C_{3D}\left( \mathbf{H}_N \right)
	=
	3
	N^2
	\cdot
	C_{1D}\left( \mathbf{H}_N \right).
	\label{eq_3d_complexity}
\end{align}
To obtain the 3D~DHT,
an overhead of $3 N^3$ additions
are necessary.
Thus,
the total additive complexity is given by:
\begin{align}
	A_{3D}\left( \mathbf{H}_N \right)
	=
	3
	N^2
	\cdot
	A_{1D}\left( \mathbf{H}_N \right)
	+
	3 N^3.
	\label{eq_3d_complexity_add}
\end{align}

We applied the obtained matrices in Section~\ref{subsec_1d_dht_approx}
in~\eqref{eq_3d_direct_approx_sym}.
A set of
3D~DHT approximations was obtained based on
such matrices and the described formalism.
Table~\ref{tab_complexity}
displays the total arithmetic complexity for
each of the
proposed
$8\times 8 \times 8$
3D~DHT approximations.
To calculate the 3D~complexity,
we applied the 1D~complexity to~\eqref{eq_3d_complexity}
and \eqref{eq_3d_complexity_add}
to obtain the approximate DHT from the approximate SDHT.
For comparison,
we included the arithmetic cost of
the exact 3D~DHT computation
according to the row-column approach applied to fast algorithm
in~\figurename~\ref{fig_fast_alg}
and
to the split radix 3D~DHT algorithm in~\cite{boussakta2001radix}.
We also included
the complexity of computing the
widely employed 3D~DCT~\cite{sawant2011balanced,li2013visualtracking,chan1997variable,Servais1997,li2007multiview,bozinovic2003scan,bozinovic2005,coutinho2017low, zaharia2002adaptive,jacob2015fpga,mulla2014image}
using the row-olumn approach
applied to the Loeffler 1D~DCT algorithm~\cite{Loeffler1989},
which presents the optimum multiplicative complexity
of 11 multiplications~\cite{Heideman1988} plus 29 additions.
Since the 3D~DCT is a separable transform,
the 3D complexity is obtained from the 1D complexity directly by~\ref{eq_3d_complexity}.
The proposed 3D~method based on matrix $\hat{\mathbf{H}}( 1 )$
presents
equal additive complexity to the exact methods.
The additive complexities of the other proposed methods increase slightly.
However,
all proposed methods present null multiplicative complexity,
which is a significant advantage
since multiplication operations
generally
demands
more hardware and energy resources~\cite{Blahut2010,oliveira2017jpeg,Liang2001}.

\begin{table}[t]
	\centering
	\caption{3D~computational complexity assessment}
	\label{tab_complexity}
	\begin{tabular}{l | ccc}
		\toprule
		&

		\multicolumn{3}{c}{3D~complexity}
		\\
		\cmidrule {2-4}
		Method & Mult. & Add. & Shift
		\\
		\midrule
		3D~DCT row-column on \cite{Loeffler1989} &
		2112 & 5568 & 0
		\\
		3D~DHT row-column on \figurename~\ref{fig_fast_alg} &
		384 & 5760 & 0
		\\
		3D~DHT by split-radix in~\cite{boussakta2001radix}  &
		384 & 5760 & 0
		\\
		Proposed 3D~method based on $\hat{\mathbf{H}}\left(1 \right)$ &
		0 & 5760 & 0 \\
		Proposed 3D~method based on $\hat{\mathbf{H}}\left(\frac{11}{8} \right)$ &
		0 & 6528 & 768 \\
		Proposed 3D~method based on $\hat{\mathbf{H}}\left(\frac{3}{2} \right)$ &
		0 & 6144 & 384 \\
		Proposed 3D~method based on $\hat{\mathbf{H}}\left(2 \right)$ &
		0 & 5760 & 384 \\
		\bottomrule
	\end{tabular}
\end{table}

\section{DICOM Data Compression Simulation}
\label{sec_dicom}

\begin{figure*}[h]
	\centering
	\begin{subfigure}[b]{.85\linewidth}
		\includegraphics[scale=1.0]{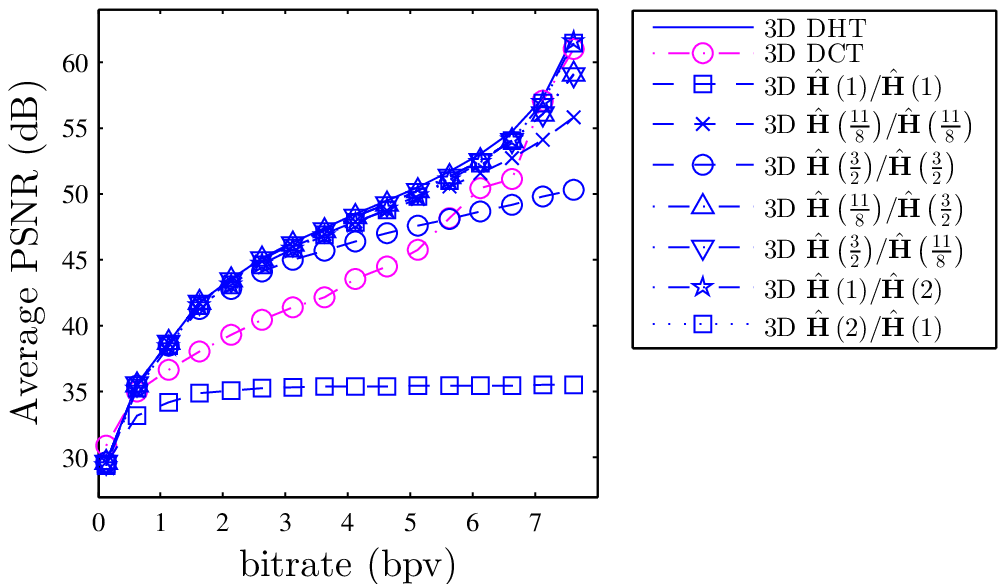}
		\caption{PSNR}
		\label{fig_psnr_xa}
	\end{subfigure} \\
	\begin{subfigure}[b]{.87\linewidth}
		\includegraphics[scale=1.0]{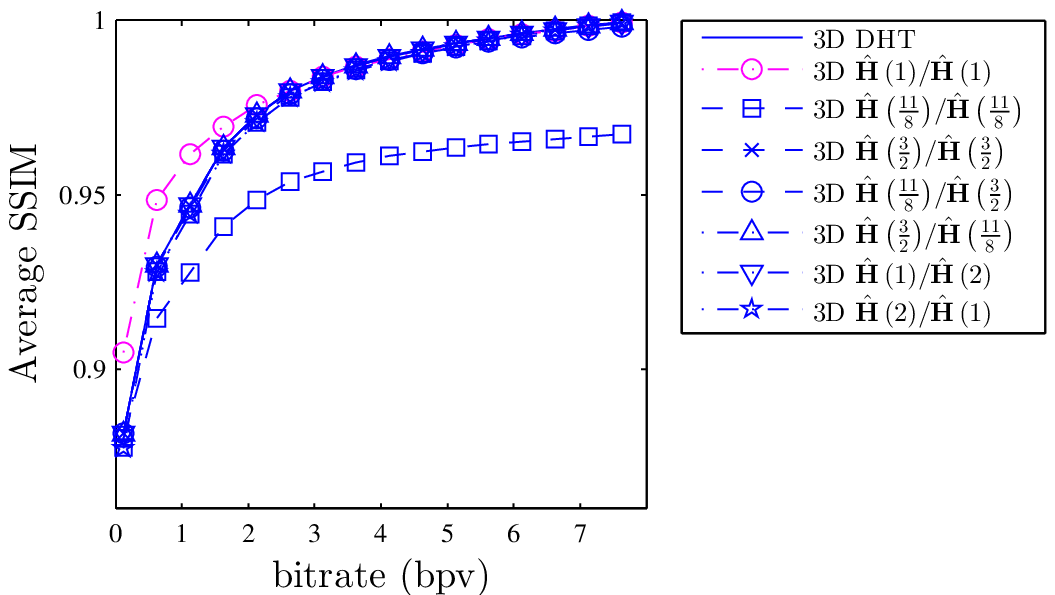}
		\caption{SSIM}
		\label{fig_ssim_xa}
	\end{subfigure}
	\caption{
		Average PSNR and SSIM versus bitrate of the 3D~DCT, the 3D~DHT, and the proposed methods for XA data.
		}
	\label{fig_measure_xa}
\end{figure*}

\begin{figure*}[h]
	\centering
	\begin{subfigure}[b]{.85\linewidth}
		\includegraphics[scale=1.0]{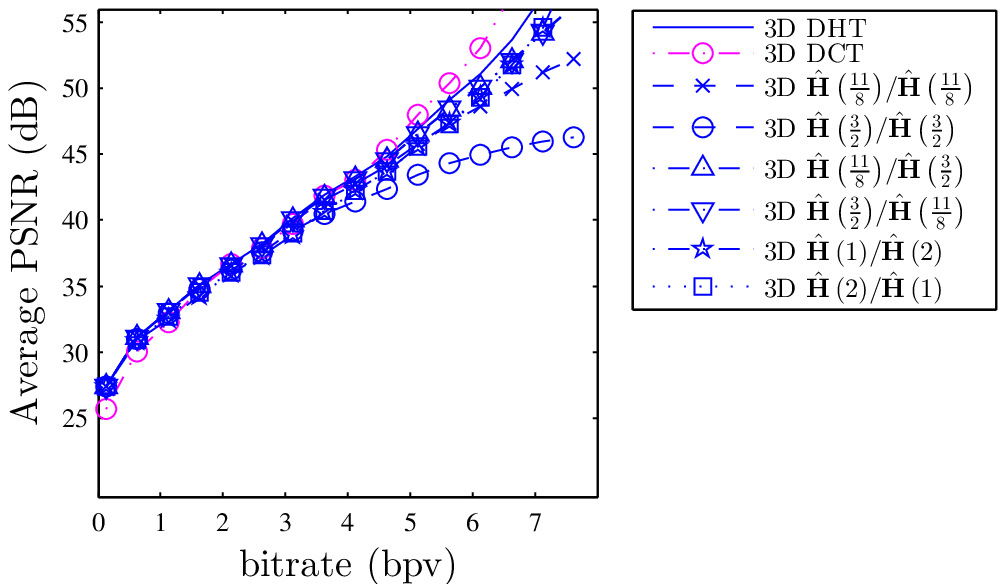}
		\caption{PSNR}
		\label{fig_psnr_mr}
	\end{subfigure} \\
	\begin{subfigure}[b]{.87\linewidth}
		\includegraphics[scale=1.0]{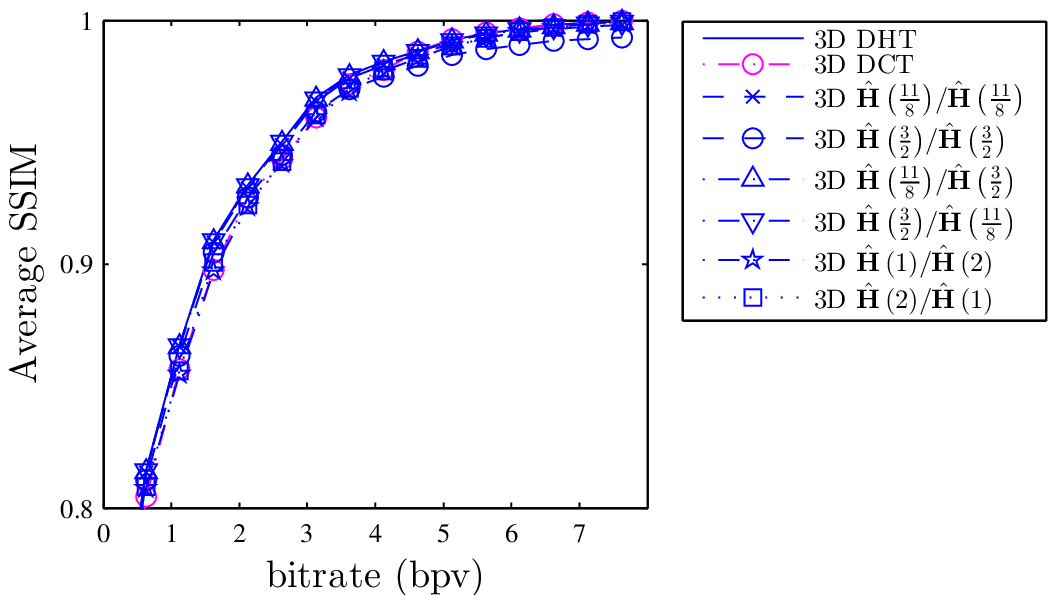}
		\caption{SSIM}
		\label{fig_ssim_mr}
	\end{subfigure}
	\caption{
		Average PSNR and SSIM versus bitrate of the 3D~DCT, the 3D~DHT, and the proposed methods for MR data.
		}
	\label{fig_measure_mr}
\end{figure*}

\begin{figure*}[t]
	\centering
	\begin{subfigure}[b]{.45\linewidth}
		\includegraphics[scale=0.9]{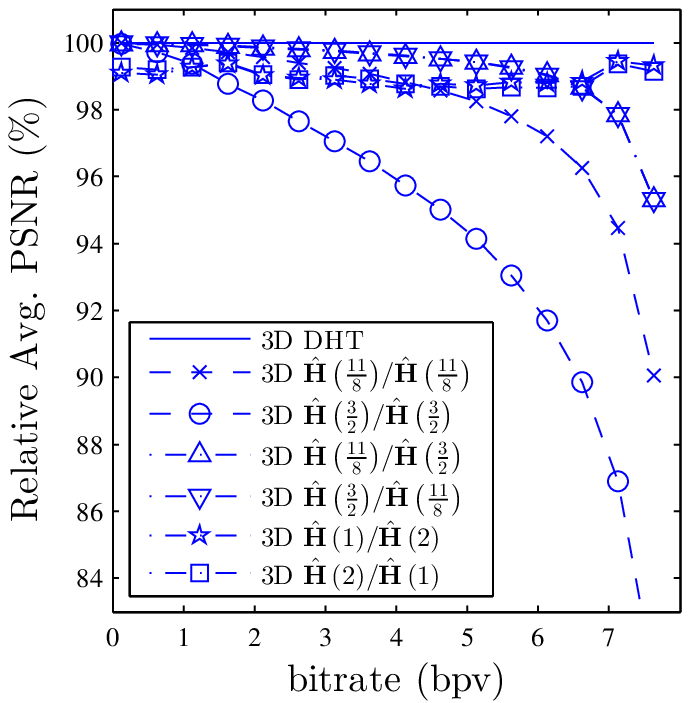}
		\caption{PSNR}
		\label{fig_rel_psnr_xa}
	\end{subfigure}
	\begin{subfigure}[b]{.45\linewidth}
		\includegraphics[scale=0.9]{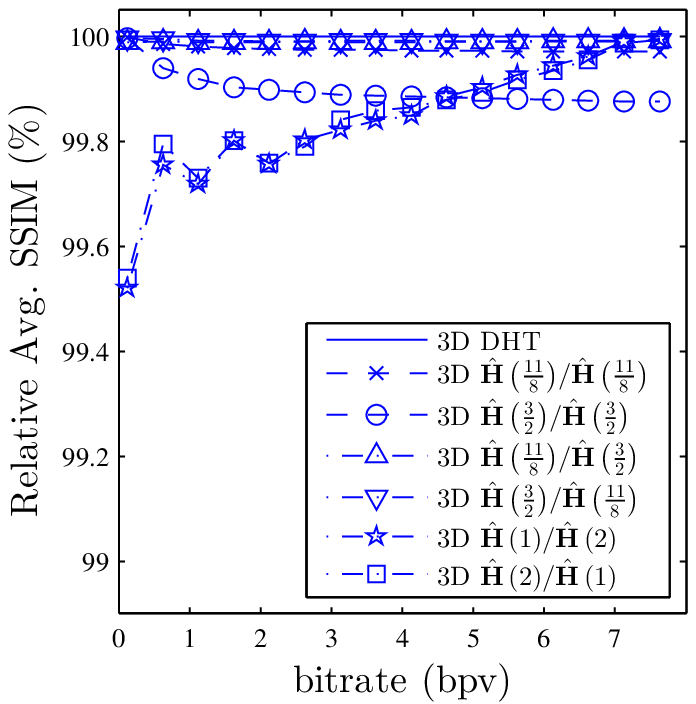}
		\caption{SSIM}
		\label{fig_rel_ssim_xa}
	\end{subfigure}
	\caption{Relative PSNR and SSIM versus bitrate of the proposed methods related to the exact 3D DHT for XA arrays.}
	\label{fig_rel_measure_xa}
\end{figure*}

\begin{figure*}[t]
	\centering
	\begin{subfigure}[b]{.45\linewidth}
		\includegraphics[scale=0.9]{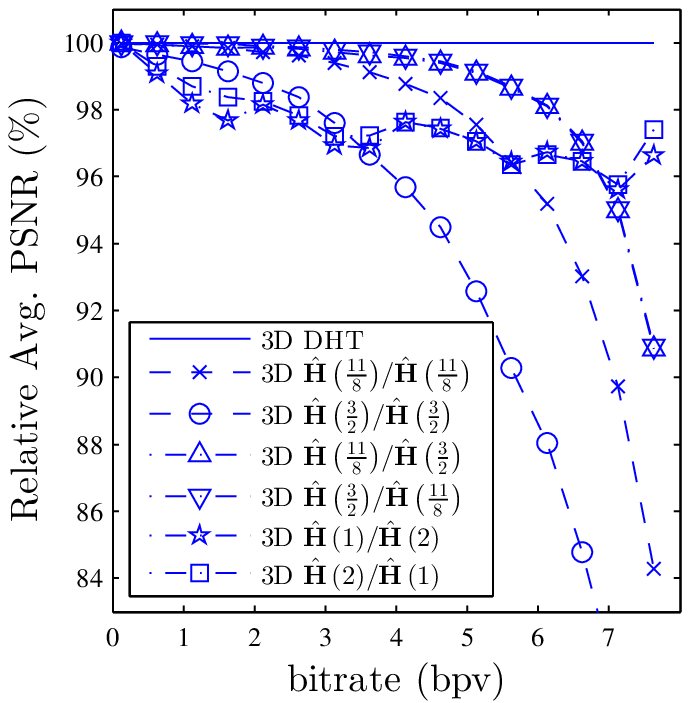}
		\caption{PSNR}
		\label{fig_rel_psnr_mr}
	\end{subfigure}
	\begin{subfigure}[b]{.45\linewidth}
		\includegraphics[scale=0.9]{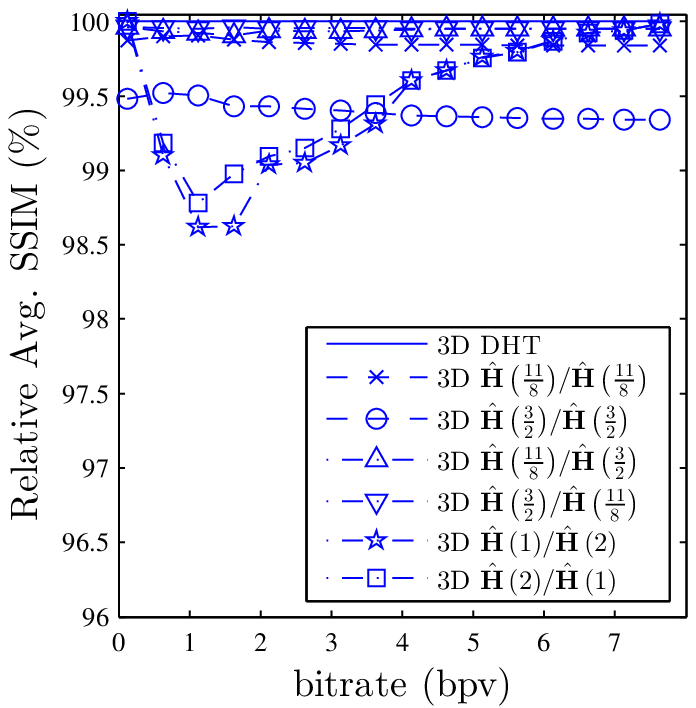}
		\caption{SSIM}
		\label{fig_rel_ssim_mr}
	\end{subfigure}
	\caption{Relative PSNR and SSIM versus bitrate of the proposed methods related to the exact 3D DHT for MR arrays.}
	\label{fig_rel_measure_mr}
\end{figure*}

\begin{figure*}[h!]
	\centering
	\begin{subfigure}[t]{0.31\linewidth}
		\includegraphics[scale=0.400]{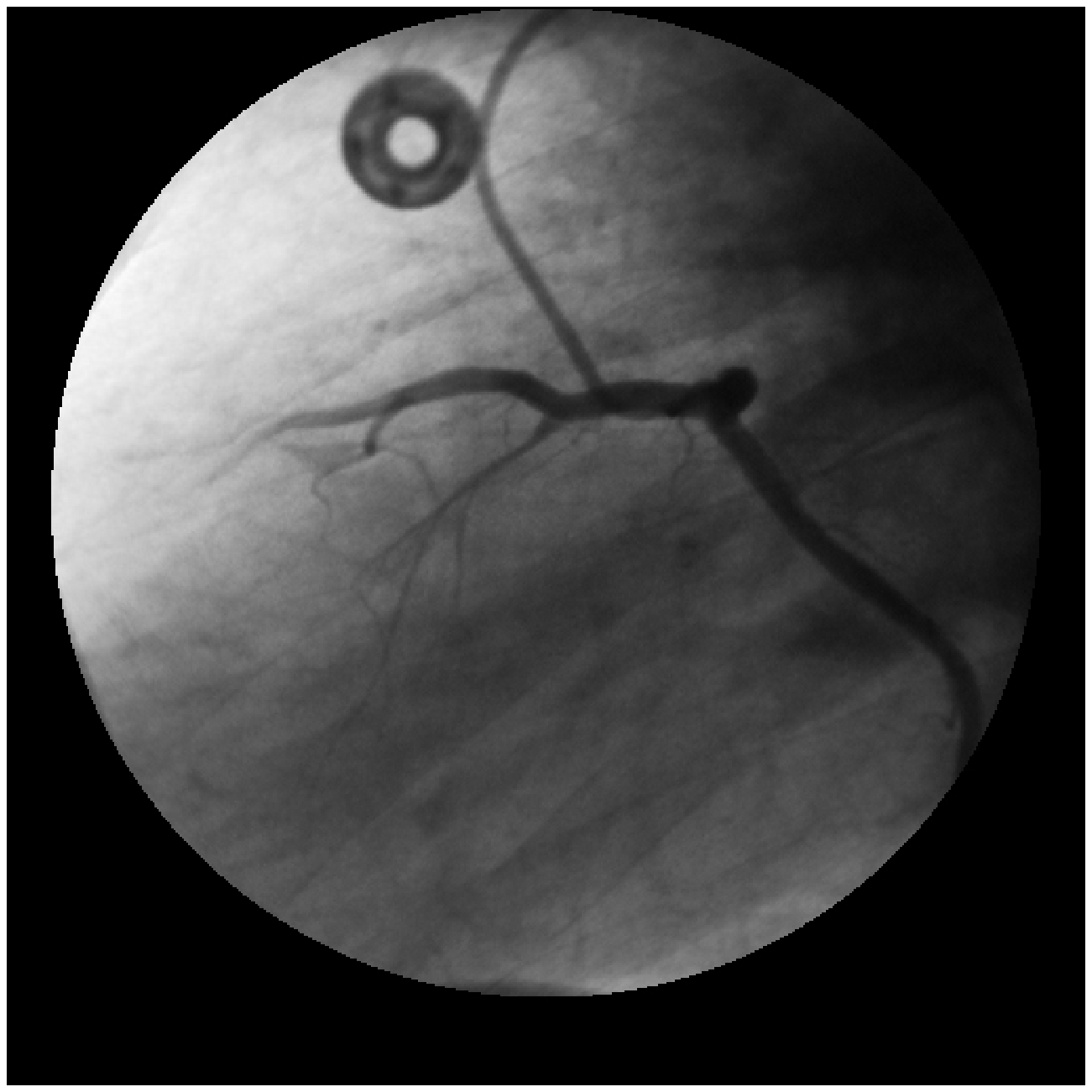}
		\caption{Uncompressed XA slice}
	\end{subfigure}
	\begin{subfigure}[t]{0.31\linewidth}
	\includegraphics[scale=0.400]{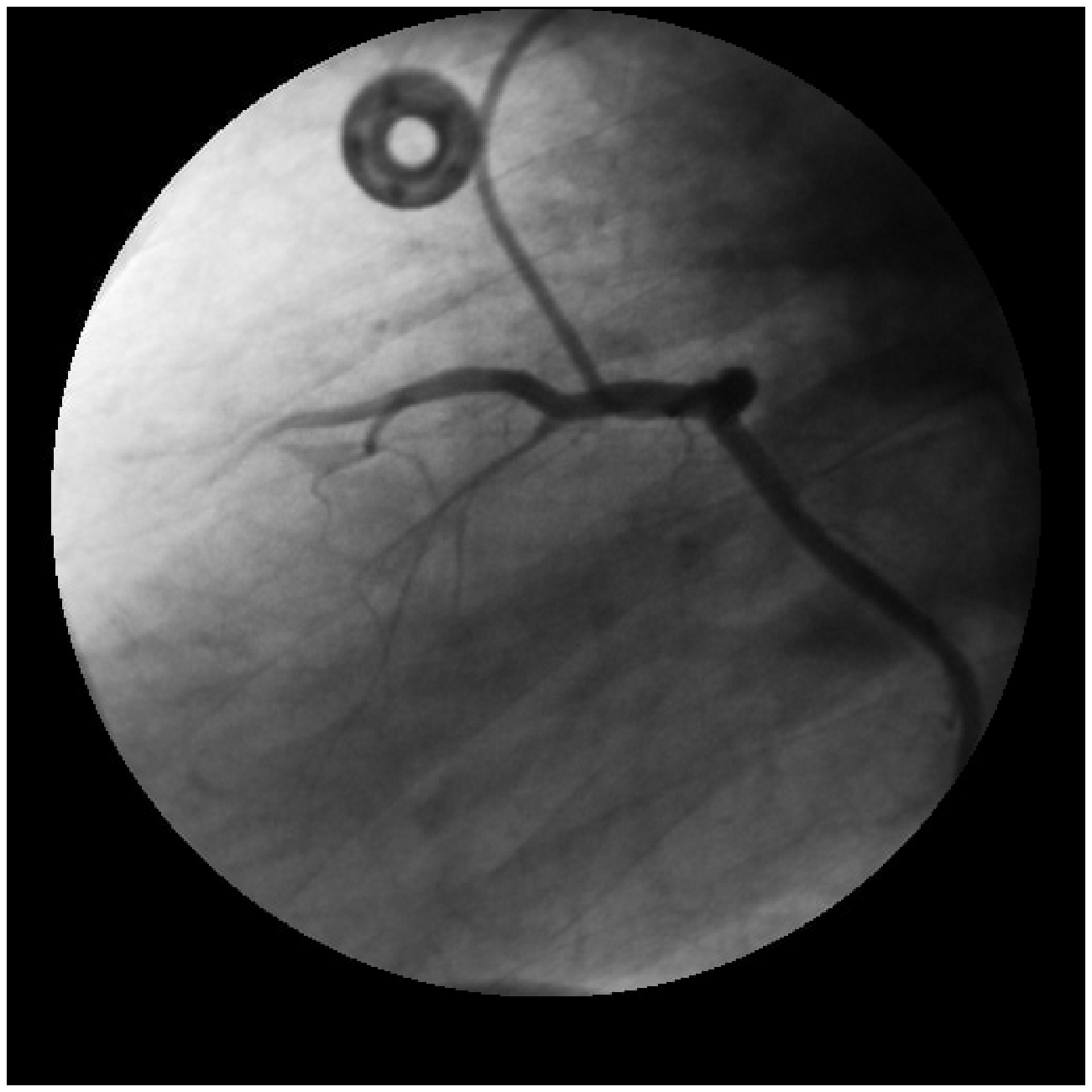}
	\caption{
		3D~DCT (PSNR=43.58~dB, SSIM=0.9883)
			}
\end{subfigure}
	\begin{subfigure}[t]{0.31\linewidth}
		\includegraphics[scale=0.400]{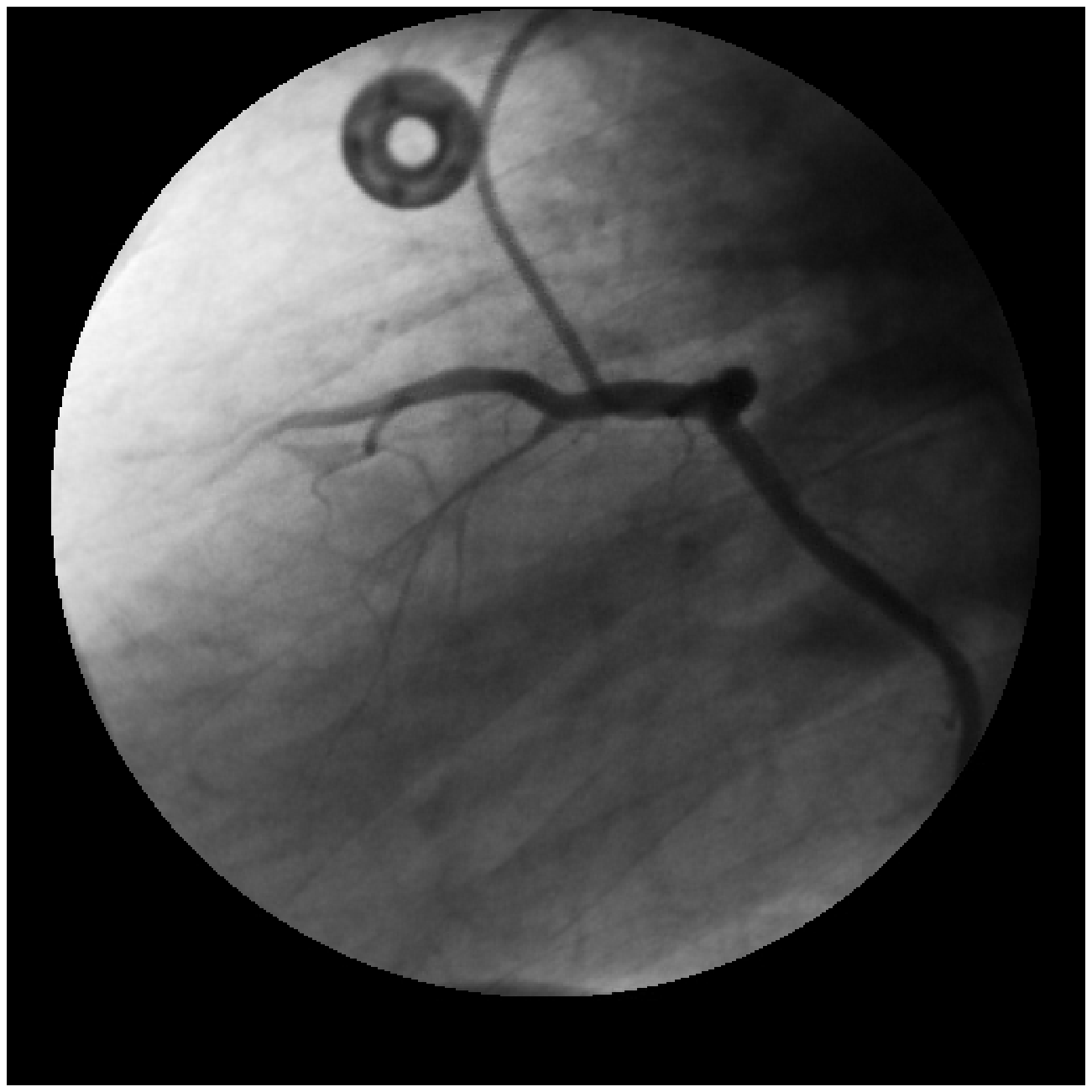}
		\caption{3D~DHT (PSNR=48.53~dB, SSIM=0.9902)}
	\end{subfigure}
	\begin{subfigure}[t]{0.31\linewidth}
		\includegraphics[scale=0.400]{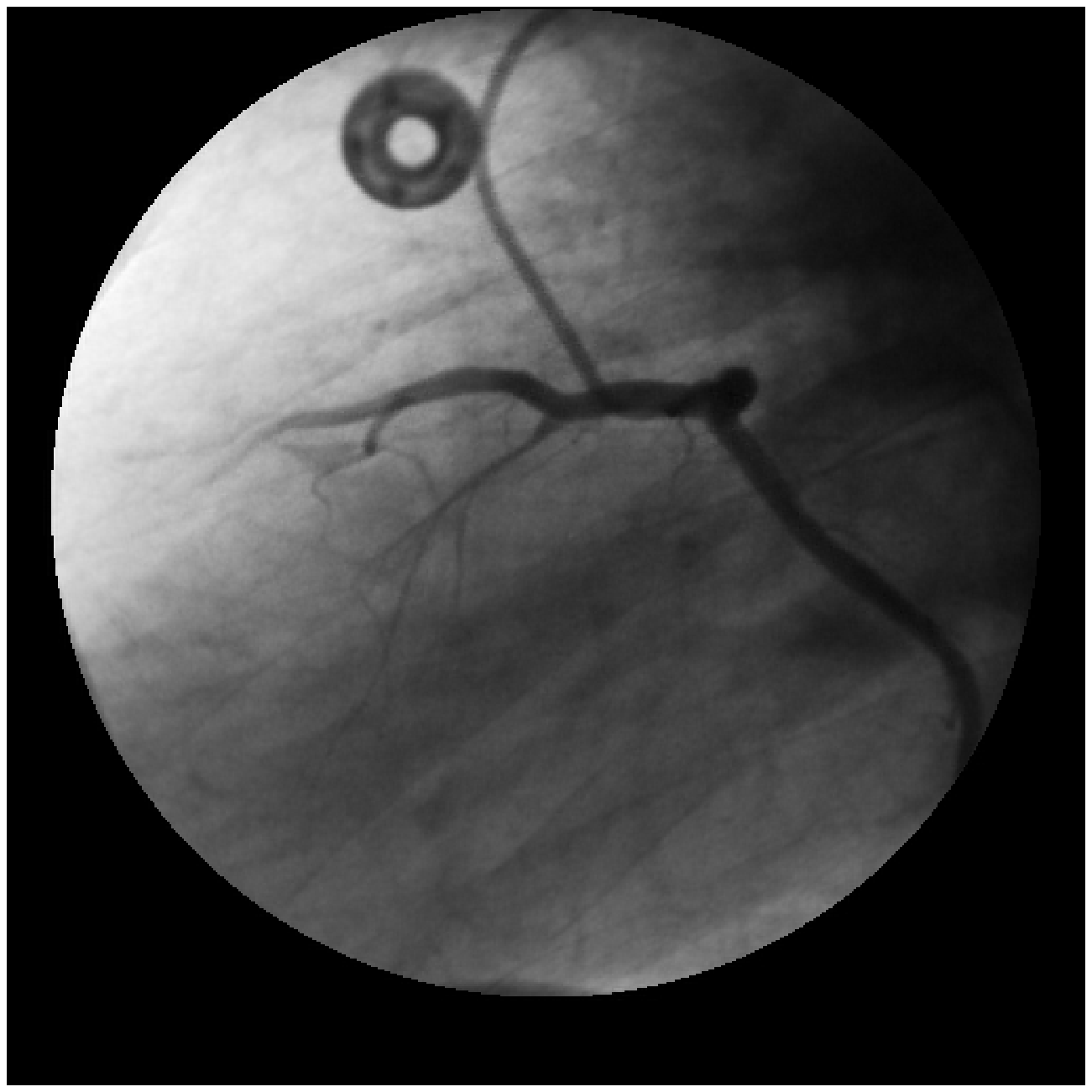}
		\caption{3D~$\hat{\mathbf{H}}\left(\frac{11}{8} \right) / \hat{\mathbf{H}}\left(\frac{11}{8} \right)$ (PSNR=47.35~dB , SSIM=0.9899)}
	\end{subfigure}
	\begin{subfigure}[t]{0.31\linewidth}
		\includegraphics[scale=0.400]{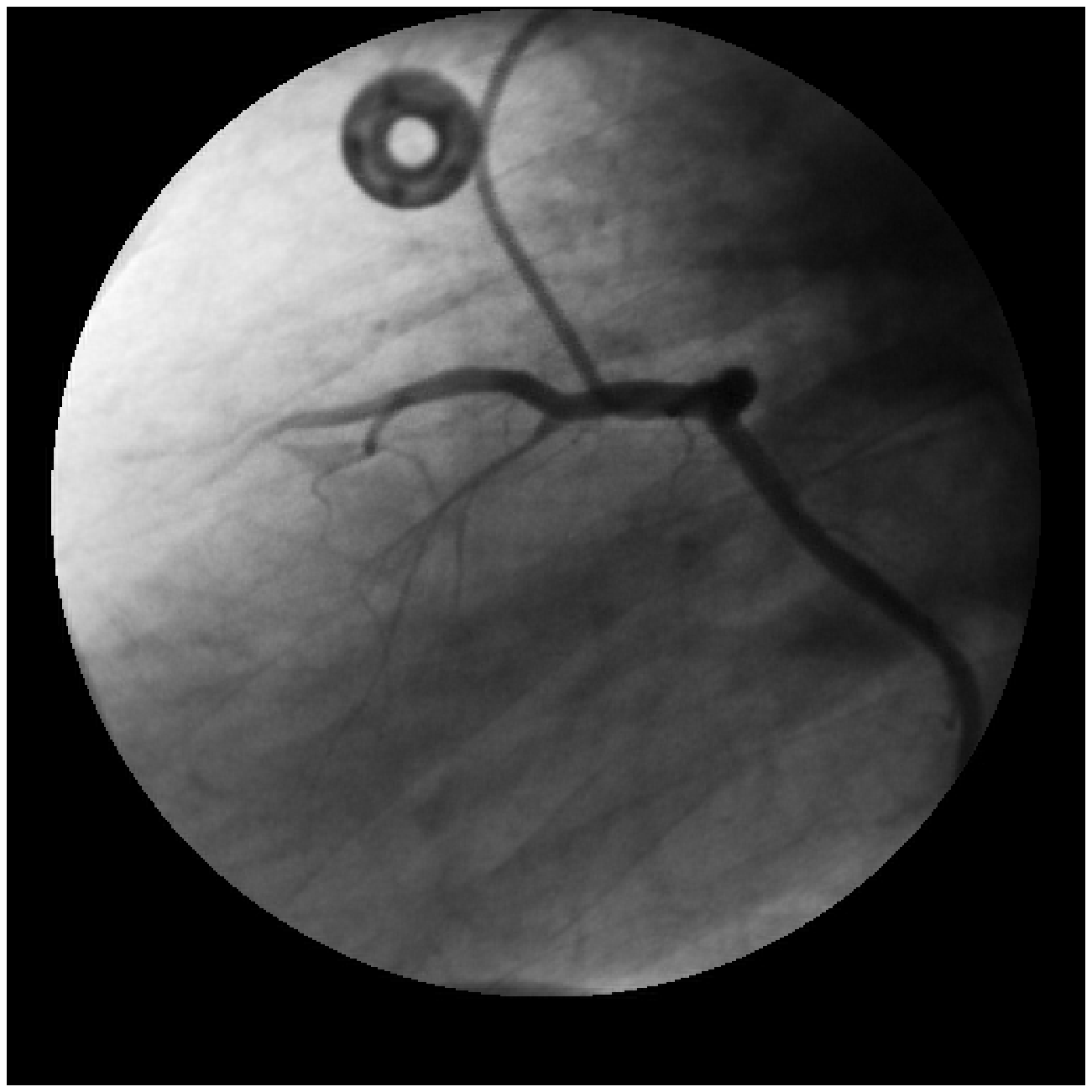}
		\caption{3D~$\hat{\mathbf{H}}\left(\frac{3}{2} \right) / \hat{\mathbf{H}}\left(\frac{3}{2} \right)$ (PSNR= 46.02~dB, SSIM=0.9889)}
	\end{subfigure}
	\begin{subfigure}[t]{0.31\linewidth}
		\includegraphics[scale=0.400]{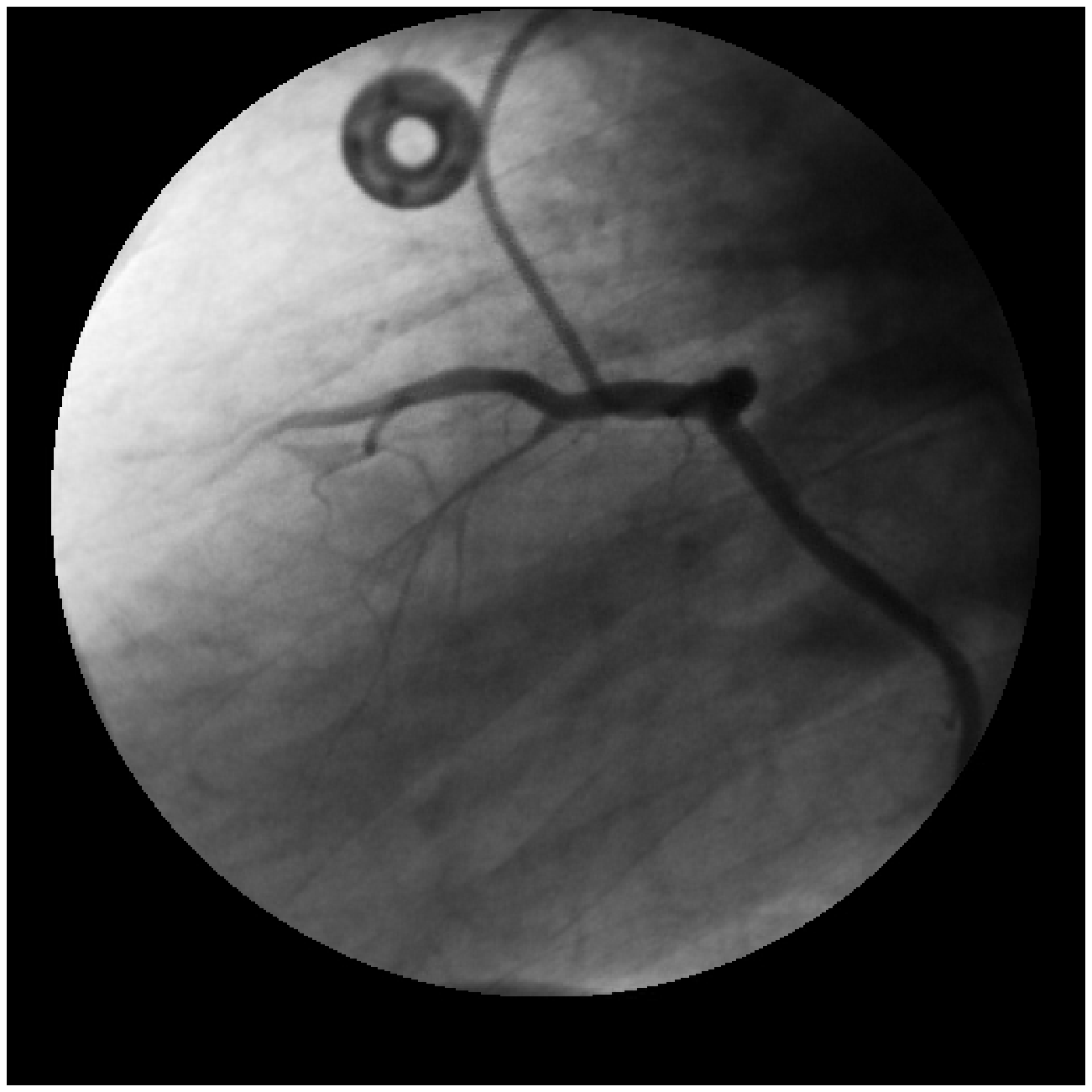}
		\caption{3D~$\hat{\mathbf{H}}\left(\frac{3}{2} \right) / \hat{\mathbf{H}}\left(\frac{11}{8} \right)$ (PSNR=48.31~dB, SSIM=0.9901)}
	\end{subfigure}
	\begin{subfigure}[t]{0.31\linewidth}
		\includegraphics[scale=0.400]{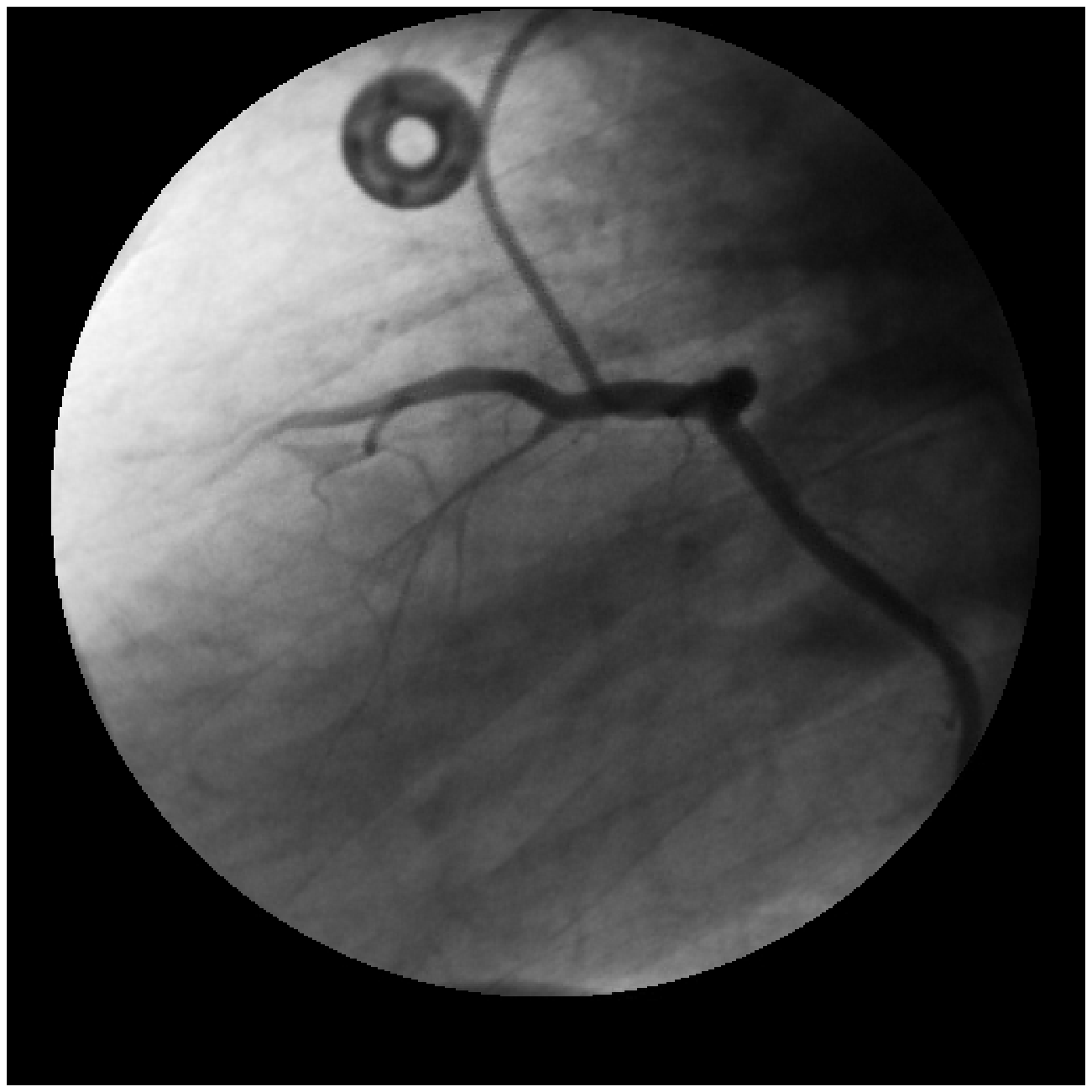}
		\caption{3D~$\hat{\mathbf{H}}\left(2 \right) / \hat{\mathbf{H}}\left(1 \right)$ (PSNR=48.04~dB, SSIM=0.9892)}
	\end{subfigure}
	\caption{Qualitative assessment for XA slice at 4 bpv compression rate.}
	\label{fig_qualitative_xa}
\end{figure*}

We employed the proposed 3D~DHT approximations
in the 3D~DHT-based medical image compression algorithm proposed in~\cite{sunder2006medical}.
We selected ten X-ray angiograms~(XA) and ten magnetic resonance~(MR)
available in~\cite{osirixdatabase,rubodatabase,zenodomr,barredatabase}
with varied resolutions of $256 \times 256$ or $512 \times 512$ and with varied number of frames/slices.
Each XA or MR data can be represented by a three-dimensional array,
i.e.,
a third-order tensor as described in Section~\ref{subsec_3d_dht}.
Each third-order tensor was divided into
smaller blocks of size $8 \times 8 \times 8$
and then
submitted to
the 3D transformation
as defined in~\eqref{eq_3d_direct_approx_sym} and~\eqref{eq_dht_from_sdht2}.
Instead of considering the zigzag scheme defined in~\cite{sunder2006medical},
in which a 2D DHT-based zigzag pattern is considered for each frame of the 3D transformed block---
we have considered a truly 3D experimental zigzag pattern,
which was obtained by averaging the amount of energy of the transform domain signal
concentrated in each coefficient and, thus,
ordering in a monotonic decreasing fashion.
The transform-domain block was
then applied to
the zigzag scheme as explained above.
For comparison,
we employed the 3D~DCT~\cite{Servais1997}
in our simulation.
Particularly for such case,
we considered the usual zigzag pattern
for the 3D~DCT~\cite{bozinovic2003scan}.

Subsequently
the data was compressed
at a fixed bitrate~\cite{bhaskaran1997,bas2008}
by preserving only
a constant number $L$ of the first zig-zag coefficients which contain the most part of signal energy.
In the full 3D~DHT-based codec implementation,
only these $L$ preserved coefficients need to be encoded by the entropy coding~\cite{Rao2001},
such as Huffman coding algorithm,
and then stored and/or transmitted.
However,
because the goal of our work is to analyze the effect of the 3D~transform block
and also because the entropy algorithms are lossless --- do not
interfere in image visual quality ---
we did not consider entropy coding in the present experiment.
The bitrate is given in \emph{bit per voxel}~(bpv)
and
computed according to
$
\operatorname{bitrate}
=
{T_b}/{T_v}
,
$~\cite{sunder2006medical},
where $T_b$ is the size of the compressed image in total of bits
and
$T_v$ is the total of voxels.
For $8 \times 8 \times 8$ blocks,
$T_v = 512$.
The fixed number $L$ of retained coefficients determines the fixed bitrate for the present compression procedure.
For instance, considering 8-bits DICOM data, to preserve 128 coefficients results in bitrate of $T_b / T_v = (8 \times 128) / 512 = 2$ bpv.

By varying the amount $L$ of coefficients retained,
we ranged the bitrate for each image
in the interval
$\left[0.125, 7.125 \right]$
with step of $0.5$.
The data compression procedure in our experiment for a fixed bitrate is described as follows:
\begin{enumerate}[\textbf{Coding step}~1:]
	\item divide the 3D~data into $B$ blocks of size $8 \times 8 \times 8$;
	\item compute the 3D~transform of each $8 \times 8 \times 8$ block, resulting in $B$ transform-domain blocks of the same size;
	\item vectorize each transform domain-block employing the 3D~zigzag pattern, resulting in $B$ zigzag vector of length~$8^3=512$;
	\item Keep only the first $L$ coefficients of each zigzag vector, discarding the remaining $512-L$.
\end{enumerate}

The data reconstruction procedure in our experiment for a fixed bitrate is described as follows:
\begin{enumerate}[\textbf{Decoding step}~1:]
	\item Reconstruct $B$ 512-length vectors employing the first $L$ preserved coefficients and filling the last $512-L$ coefficients with zeros;
	\item Employ the inverse 3D~zigzag scheme to restore $B$ $8 \times 8 \times 8$ compressed transform-domain blocks;
	\item Compute the inverse 3D~transform of each $8 \times 8 \times 8$ block, resulting in $B$ image-domain recovered blocks of the same size;
	\item Employ the $B$ recovered image-domain blocks of size $8 \times 8 \times 8$ to reconstruct the compressed 3D~data.
\end{enumerate}

As figure of merit measures for performance evaluation,
it was utilized the \emph{peak signal-to-noise ratio}~(PSNR)~\cite[p.~9]{bhaskaran1997}
and the \emph{structural similarity index}~(SSIM)~\cite{Wang2004}.
The average between each metric obtained from each array
was considered and weighted according to the number of blocks of each array.
Our simulations showed that the discussed
methods offer a similar behavior,
except for the involutional method $\hat{\mathbf{H}}\left(1 \right)/\hat{\mathbf{H}}\left(1 \right)$.
In fact,
such method
generates the highest deviation from diagonality,
as shown in Table~\ref{tab_metrics},
which introduces large error when computing the inverse transformation.
Thus, we excluded this case from our analysis.
It is presented in \figurename~\ref{fig_measure_xa} and \figurename~\ref{fig_measure_mr} the average result for XA and MR, respectively, for all the considered methods and different bitrates.
These results are also presented in \tablename~\ref{tab_performance_xa} and in \tablename~\ref{tab_performance_mr}.

\begin{table}
	\centering
	\caption{Average PSNR~(dB) and SSIM measures (respectively) for each method for different bitrates for XA datas}
	\label{tab_performance_xa}
	\begin{tabular}{l | c c c c c c c c}
		\toprule
		 & \multicolumn{8}{c}{Bitrate} \\
		 \cmidrule{2-9}
		 Method & 0.125 & 1.125 & 2.125 & 3.125 & 4.125 & 5.125 & 6.125 & 7.125 \\
		\midrule
		\multirow{2}{*}{3D~DHT} &  29.62 & 38.73 & 43.52 & 46.33 & 48.46 & 50.52 & 53.05 & 57.28 \\
		& .8815 &  .9471 &  .9730 &  .9840 &  .9898 &  .9935 &  .9963 &  .9986 \\
		\midrule
		\multirow{2}{*}{3D~DCT} & 30.89 & 36.64 & 39.30 & 41.36 & 43.56 & 45.77 & 50.40 & 57.04 \\
		& .9047 & .9616 & .9758  & .9838 & .9890 & .9929  & .9960  & .9984 \\
		\midrule
		\multirow{2}{*}{$\hat{\mathbf{H}}\left(\frac{11}{8} \right) / \hat{\mathbf{H}}\left(\frac{11}{8} \right)$}
		& 29.62 & 38.68 & 43.34 & 45.98 & 47.90 & 49.63 & 51.56 & 54.11 \\
		& .8815 &  .9469 &  .9728 &  .9837 &  .9895 &  .9932 &  .9960 &  .9983 \\
		\midrule
		\multirow{2}{*}{$\hat{\mathbf{H}}\left(\frac{3}{2} \right) / \hat{\mathbf{H}}\left(\frac{3}{2} \right)$}
		& 29.61 & 38.49 & 42.77 & 44.97 & 46.40 & 47.56 & 48.65 & 49.77 \\
		& .8815 &  .9463 &  .9720 &  .9829 &  .9887 &  .9923 &  .9951 &  .9974 \\
		\midrule
		\multirow{2}{*}{$\hat{\mathbf{H}}\left(\frac{11}{8} \right) / \hat{\mathbf{H}}\left(\frac{3}{2} \right)$}
		& 29.61 & 38.71 & 43.46 & 46.21 & 48.27 & 50.23 & 52.54 & 56.05 \\
		& .8814 &  .9470 &  .9729 &  .9838 &  .9897 &  .9934 &  .9962 &  .9985 \\
		\midrule
		\multirow{2}{*}{$\hat{\mathbf{H}}\left(\frac{3}{2} \right) / \hat{\mathbf{H}}\left(\frac{11}{8} \right)$}
		& 29.62 & 38.71 & 43.46 & 46.22 & 48.28 & 50.23 & 52.53 & 56.05 \\
		& .8815 &  .9470 &  .9729 &  .9839 &  .9897 &  .9934 &  .9962 &  .9985 \\
		\midrule
		\multirow{2}{*}{$\hat{\mathbf{H}}\left(1 \right) / \hat{\mathbf{H}}\left(2 \right)$}
		& 29.35 & 38.44 & 43.10 & 45.81 & 47.79 & 49.88 & 52.41 & 56.96 \\
		&  .8773 &  .9444 &  .9706 &  .9822 &  .9883 &  .9925 &  .9957 &  .9985 \\
		\midrule
		\multirow{2}{*}{$\hat{\mathbf{H}}\left(2 \right) / \hat{\mathbf{H}}\left(1 \right)$}
		& 29.40 & 38.45 & 43.11 & 45.87 & 47.87 & 49.83 & 52.33 & 56.92 \\
		&  .8775 &  .9446 &  .9707 &  .9824 &  .9885 &  .9924 &  .9956 &  .9985 \\
		\bottomrule
	\end{tabular}
\end{table}

\begin{table}
	\centering
	\caption{Average PSNR~(dB) and SSIM measures (respectively) for each method for different bitrates for MR datas}
	\label{tab_performance_mr}
	\begin{tabular}{l | c c c c c c c c}
		\toprule
		& \multicolumn{8}{c}{Bitrate} \\
		\cmidrule{2-9}
		Method & 0.125 & 1.125 & 2.125 & 3.125 & 4.125 & 5.125 & 6.125 & 7.125 \\
		\midrule
		\multirow{2}{*}{3D~DHT} &  27.40 & 33.13 & 36.68 & 40.15 & 43.30 & 46.95 & 51.07 & 57.07 \\
		& .7145 &  .8668 &  .9326 &  .9682 &  .9832 &  .9920 &  .9966 &  .9990 \\
		\midrule
		\multirow{2}{*}{3D~DCT} &  25.67 & 32.23 & 36.70 & 39.72 & 43.06 & 48.03 & 53.04 & 61.09 \\
		& .6985 &  .8568 &  .9274 &  .9600 &  .9798 &  .9921 &  .9969 &  .9992 \\
		\midrule
		\multirow{2}{*}{$\hat{\mathbf{H}}\left(\frac{11}{8} \right) / \hat{\mathbf{H}}\left(\frac{11}{8} \right)$}
		& 27.39 & 33.09 & 36.57 & 39.91 & 42.77 & 45.80 & 48.61 & 51.21 \\
		& .7136 &  .8660 &  .9313 &  .9668 &  .9816 &  .9905 &  .9950 &  .9974 \\
		\midrule
		\multirow{2}{*}{$\hat{\mathbf{H}}\left(\frac{3}{2} \right) / \hat{\mathbf{H}}\left(\frac{3}{2} \right)$}
		& 27.36 & 32.94 & 36.24 & 39.19 & 41.43 & 43.46 & 44.96 & 45.99 \\
		& .7108 &  .8625 &  .9273 &  .9624 &  .9770 &  .9856 &  .9901 &  .9924 \\
		\midrule
		\multirow{2}{*}{$\hat{\mathbf{H}}\left(\frac{11}{8} \right) / \hat{\mathbf{H}}\left(\frac{3}{2} \right)$}
		& 27.40 & 33.09 & 36.63 & 40.03 & 43.11 & 46.53 & 50.10 & 54.21 \\
		& .7142 &  .8662 &  .9320 &  .9676 &  .9826 &  .9915 &  .9961 &  .9985 \\
		\midrule
		\multirow{2}{*}{$\hat{\mathbf{H}}\left(\frac{3}{2} \right) / \hat{\mathbf{H}}\left(\frac{11}{8} \right)$}
		& 27.40 & 33.12 & 36.64 & 40.07 & 43.12 & 46.55 & 50.12 & 54.25 \\
		& .7142 &  .8664 &  .9321 &  .9678 &  .9827 &  .9915 &  .9961 &  .9985 \\
		\midrule
		\multirow{2}{*}{$\hat{\mathbf{H}}\left(1 \right) / \hat{\mathbf{H}}\left(2 \right)$}
		& 27.40 & 32.52 & 36.00 & 38.92 & 42.27 & 45.57 & 49.39 & 54.57 \\
		&  .7145 &  .8549 &  .9236 &  .9602 &  .9793 &  .9896 &  .9953 &  .9984 \\
		\midrule
		\multirow{2}{*}{$\hat{\mathbf{H}}\left(2 \right) / \hat{\mathbf{H}}\left(1 \right)$}
		& 27.40 & 32.70 & 36.04 & 39.03 & 42.27 & 45.57 & 49.36 & 54.66 \\
		& .7145 &  .8563 &  .9242 &  .9612 &  .9793 &  .9896 &  .9953 &  .9984 \\
		\bottomrule
	\end{tabular}
\end{table}

Generally,
all the methods present better performance at compressing XA data than MR data.
For such case,
the exact 3D~DHT and the proposed
approximate versions present better performance than the 3D~DCT
in terms of PSNR for the bitrate ranging from 1 bpv to 6 bpv.
For instance,
at 2 bpv,
the exact 3D~DHT and the proposed 3D~$\hat{\mathbf{H}}\left(\frac{3}{2} \right) / \hat{\mathbf{H}}\left(\frac{11}{8} \right)$
present 10.7 \% and 10.6 \% higher PSNR than the 3D~DCT, respectively;
for 6 bpv,
such values are 5.2 \% and 4.2\%, respectively.
In terms of SSIM,
the 3D~DCT outperform the 3D~DHT and the proposed methods
in the shorter range of (0~bpv, 2~bpv).
For instance,
at 2 bpv, the exact 3D~DHT and the proposed
$\hat{\mathbf{H}}\left(\frac{3}{2} \right) / \hat{\mathbf{H}}\left(\frac{11}{8} \right)$
present 0.28 \% and 0.23 \% lower SSIM than the 3D~DCT, respectively;
whereas for 6 bpv they present
0.03 \% and 0.02 \% higher SSIM than the 3D~DCT, respectively.
The results for compressing MR data are very close,
with the 3D~DCT presenting slightly lower PSNR values for low bitrates (<2 bpv)
and higher values for high bitrates (>5 bpv).
For example,
at 1 bpv, the
exact 3D~DHT and the proposed
$\hat{\mathbf{H}}\left(\frac{3}{2} \right) / \hat{\mathbf{H}}\left(\frac{11}{8} \right)$
present 2.8 \% and 2.8 \% higher PSNR than the 3D~DCT,
whereas, for 6 bpv,
they present 3.7 \% and 5.5 \% lower PSNR than the 3D~DCT.
Generally,
all the methods present a competitive performance compared to the 3D~DCT.
However,
the exact 3D~DHT shows a substantial reduction in the multiplicative complexity of
of 81.8 \% (see \tablename~\ref{tab_complexity}),
while all the proposed 3D~DHT approximations present
100 \% of multiplicative complexity reduction.

Hence,
we focus on comparing the proposed 3D~DHT approximations
with the exact 3D~DHT.
In \figurename~\ref{fig_rel_measure_xa} and \figurename~\ref{fig_rel_measure_mr},
it is shown a comparison of the proposed methods with the original 3D~DHT,
in which it is presented
the relative PSNR and SSIM compared to the exact 3D~DHT-based compression.

For low bitrates ($<2$ bpv)
the general performance is higher than
$98 \%$ in terms of PSNR and
$99 \%$ in terms of SSIM compared to the exact 3D~DHT.
For XA images, specifically,
the performance of the proposed methods is higher than $99.5 \%$ in terms of SSIM.
The simulations suggest that all methods present better performance at compressing XA than MR images.
When increasing the bitrate,
the 3D~DHT approximations relative PSNR decrease compared to the 3D~DHT.
Such percentage difference occurs because,
for the exact 3D~DHT,
the exact inverse 3D~DHT is employed to reconstruct the image
and thus it achieves perfect reconstruction when the bitrate $\rightarrow 8$~bpv,
which results in PSNR $\rightarrow \infty$.
The proposed 3D~DHT approximations,
however,
consider low-complexity approximate versions for the their inverse 3D transformations.
Then,
they do not achieve the perfect reconstruction when $\rightarrow 8$~bpv and the relative PSNR will increase for high bitrates.
Such effect is not observed in the relative SSIM values because the SSIM measure ranges in the closed interval $[0,1]$.
Instead,
the relative SSIM increases with the bitrate.
Furthermore,
the absolute PSNR values show that
the performance for bitrates higher than 4 bpv are higher than 40 dB for all the methods.
Thus, the relative PSNR does not capture the
actual perceived image quality difference.
As a demonstration of such aspect,
a qualitative comparison is shown in
\figurename~\ref{fig_qualitative_xa}
for a slice of
an XA compressed at 4~bpv employing some of the proposed methods.
The compressed images are very close to the uncompressed image
and the approximate methods performances are practically indistinguishable when compared to the exact 3D~DHT.

Such results show that our proposed approximate 3D~DHT-based methods are
as suitable to coding medical images as the
exact 3D~DHT-based method in~\cite{sunder2006medical}
at a considerable lower complexity cost.
Generally,
the proposed approximate methods present more than
$98\%$ of SSIM performance relative to the original method
at a much
lower computational cost.

\section{Time and Memory Complexity Assessment}
\label{sec_time_mem_complexity}

In the present Section,
we aim at analyzing the time and memory complexity of the proposed 3D approximate transforms in
a real-time processing scenario
and compare them with
the both traditional 3D~DHT and 3D~DCT.
To demonstrate the appropriateness of the proposed methods in
real-time embedded applications,
we implemented the proposed 3D transforms in an ARM Cortex-M0+ processor.
The ARM Cortex cores have been widely adopted in low-power and real-time embedded applications,
such as internet of things devices~\cite{hoppner2019achieve, zhang2018recryptor,onuki2017embedded}.

We employed the Raspberry Pi Pico board~\cite{rasp_pipico},
which is equipped with the recently introduced RP2040 microcontroller.
The RP2040 is a low cost and low power chip designed by Raspberry Pi,
presenting a dual-core ARM Cortex-M0+ processor with 264KB internal random access memory.
In addition,
we considered the C language with the Software Development Kit~(SDK) libraries for C/C++~\cite{rasp_pipico_sdk}
in our code implementation.

Following the mathematical development of Section~\ref{sec_3d_dht},
we implemented the 3D transforms by means of several 1D algorithms
for each vector in each dimension.
Such procedure is employed to compute the 3D~SDHT,
the proposed 3D~SDHT aproximations
and the 3D~DCT.
To compute the 1D algorithms for the DHT and the
proposed approximate versions,
the flow-graph of \figurename~\ref{fig_fast_alg}
is implemented.
For the DCT case,
we considered the
Loeffler algorithm~\cite{Loeffler1989},
which is optimum in terms of multiplicative complexity of the DCT~\cite{Heideman1988}.
The procedure is summarized as follows:

\begin{enumerate}[i)]
\item compute $64$ 1D transforms of each row (1st dimension);

\item transpose the array by shifting dimensions, i.e., $n_1 \times n_2 \times n_3$ turns to $n_2 \times n_3 \times n_1$;

\item compute $64$ 1D transforms of each row of the transposed array (2nd dimension);

\item transpose the array by shifting dimensions, i.e., $n_2 \times n_3 \times n_1$ turns to $n_3 \times n_1 \times n_2$;

\item compute $64$ 1D transforms of each row of the transposed array (3rd dimension);

\item transpose the array back to its original position by shifting dimensions, i.e., $n_3 \times n_1 \times n_2$ turns to $n_1 \times n_2 \times n_3$.

\end{enumerate}

Next,
it is required to
employ the 3D~SDHT (or an approximate version)
to compute
the
3D~DHT (or an approximate version) according to
\eqref{eq_dht_manipulation} and
\eqref{eq_dht_from_sdht2}.
Such operation requires only additions,
as described in Section~\ref{subsec_complex_assess}
and it is implemented with a triple ``for'' loop.
Notice that the 3D~DCT does not require such procedure because it is already a separable transform.

Our implementation is performed sequentially in a single core
and
considers $8 \times 8 \times 8$ input arrays of 32-bits integers.
For the exact 3D~DHT and the 3D~DCT cases,
the multiplications by the irrational quantities
are performed via floating point multiplication,
i.e.,
the input integer coefficient is converted to floating point representation (via type casting),
multiplied by the irrational constant,
and then converted back to integer.
On the other hand,
for the proposed approximate 3D~DHTs,
the multiplications by the dyadic rationals are
realized by integers bitshifts and additions only,
according to Table~\ref{tab_beta_parameters}.
Thus,
the proposed methods do not require any multiplication operation at all,
as presented in Section~\ref{subsec_complex_assess}.

We compute the 3D transforms of 8,192 blocks of size $8\times 8 \times 8$,
randomly generated.
For instance,
such number of blocks is equivalent to a full DICOM data of size
$265 \times 256 \times 64$.
Our goal is to analyze the elapsed time and the memory usage
for each 3D transform employed in the present experiment.
The time elapsed was obtained employing the RP2040 timestamp function
available in the SDK library. The results are shown in \tablename~\ref{tab_elapsed_time}.
We repeated the above experiment fifty times and considered the average time elapsed.

It can be observed that the proposed methods present a substantial
reduction in terms of execution time compared to both the 3D~DCT and the 3D~DHT.
Particularly,
the proposed 3D~method based on $\hat{\mathbf{H}}\left(1 \right)$
presented the fastest execution of 3200 ms,
which corresponds to percentage reductions compared to the 3D~DCT and 3D~DHT of 91.22~\% and 72.53~\%, respectively.
Such result is expected
since such 3D method presents the lowest arithmetic complexity,
as shown in Table~\ref{tab_complexity}.
Generally,
all the proposed 3D~methods present significant reduction in execution time,
with percentage reduction compared to the 3D~DCT and 3D~DHT of $\sim$90 \% and $\sim$70\%.
Such results are consequence of the fact the proposed methods present null multiplicative complexity.

\begin{table}
	\centering
	\caption{ Execution time for each 3D~method and the respective percentage reduction compared to the 3D~DCT execution time employing the RP2040 microcontroller}
	\label{tab_elapsed_time}
	\begin{tabular}{l | c c c}
		\toprule
		Method & Elapsed time (ms) & reduction to 3D~DCT & reduction to 3D~DHT
		\\
		\midrule
		3D~DCT row-column on \cite{Loeffler1989} &
		36457 & --- & ---
		\\
		3D~DHT row-column on \figurename~\ref{fig_fast_alg} &
		11650 & 68.04 \% & ---
		\\
		Proposed 3D~method based on $\hat{\mathbf{H}}\left(1 \right)$ &
		3200 & \textbf{91.22} \% & \textbf{72.53} \% \\
		Proposed 3D~method based on $\hat{\mathbf{H}}\left(\frac{11}{8} \right)$ &
		3637  & 90.02 \% & 68.78 \% \\
		Proposed 3D~method based on $\hat{\mathbf{H}}\left(\frac{3}{2} \right)$ &
		3513 & 90.36 \% & 69.84 \% \\
		Proposed 3D~method based on $\hat{\mathbf{H}}\left(2 \right)$ &
		3521 & 90.34 \% & 69.77 \% \\
		\bottomrule
	\end{tabular}
\end{table}

For the memory usage, we analyzed the memory usage observing the compiler output \emph{Executable and Linkable Format}~(ELF) file~\cite{arm_elf}.
We considered the \texttt{.text} segment, which is a measure of the code length~\cite[p.~64]{sytemv_abi}, and the the \texttt{.data} and the \texttt{.bss} segments, which hold the initialized and uninitialized data, respectively, and measure together the statically allocated memory~\cite[p.~62]{sytemv_abi} --- we did not employ dynamic allocation in the present experiment.
The memory usage for each method is shown in \tablename~\ref{tab_memory}.
In general,
all the methods present the same usage for statically allocated memory
and minor difference in code length.
Such result is expected since the proposed methods address
the reduction in arithmetic complexity,
which results in a faster execution but preserve
the same structure in terms of
transform and vector lengths.

\begin{table}[b]
	\centering
	\caption{Memory usage for each 3D~method employing the RP2040 microcontroller}
	\label{tab_memory}
	\begin{tabular}{l | ccc}
		\toprule
		&

		\multicolumn{2}{c}{Memory usage (bytes)}
		\\
		\cmidrule {2-3}
		Method & \texttt{.text} & \texttt{.data} + \texttt{.bss}
		\\
		\midrule
		3D~DCT row-column on \cite{Loeffler1989} &
		30160 & 1072 + 3428 = 4500
		\\
		3D~DHT row-column on \figurename~\ref{fig_fast_alg} &
		29904 & 1072 + 3428 = 4500
		\\
		Proposed 3D~method based on $\hat{\mathbf{H}}\left(1 \right)$ &
		29472 & 1072 + 3428 = 4500
		\\
		Proposed 3D~method based on $\hat{\mathbf{H}}\left(\frac{11}{8} \right)$ &
		29416 & 1072 + 3428 = 4500  \\
		Proposed 3D~method based on $\hat{\mathbf{H}}\left(\frac{3}{2} \right)$ &
		29376 & 1072 + 3428 = 4500 \\
		Proposed 3D~method based on $\hat{\mathbf{H}}\left(2 \right)$ &
		29376 & 1072 + 3428 = 4500 \\
		\bottomrule
	\end{tabular}
\end{table}

\section{Conclusion}
\label{sec_conc}

In the present work,
we proposed a mathematical formulation
for approximating the 3D~DHT
based on
tensor formalism.
A parametric exhaustive search
for deriving approximations for the DHT matrix
was performed.
A set of quasi-orthogonal approximate DHT matrices was presented.
Such matrices are optimized according to performance,
arithmetic complexity and quasi-orthogonality measures.

A collection of $8 \times 8 \times 8$ multiplierless 3D~DHT approximations based on the derived approximate matrices
was proposed.
The derived 3D~transforms were applied
to the 3D~DHT-based DICOM medical image coding scheme in~\cite{sunder2006medical}.
We also included the 3D~DCT
for comparison.
The results showed very close performance~($>98\%$ in terms of SSIM relative to the original method)
at a considerable lower computational cost ($100\%$ multiplicative complexity reduction).
The proposed methods are realistic low-complexity coding algorithms that can be employed
in medical image compression, transmission and storage systems presenting rigorous energy and hardware resources constraints.

Furthermore,
we have implemented
the proposed methods
as well as the classical 3D~DCT and 3D~DHT in an ARM Cortex-M0+ processor employing the recently introduced Raspberry Pi Pico board.
We analyzed the execution time and memory consumption.
The results show a noticeable execution time reduction of $\sim 90\%$ and $\sim 70 \%$ compared to the 3D~DCT and the 3D~DHT, respectively,
whereas the memory usage is not significantly affected.
Thus,
we conclude that the main
advantage of the proposed methods
is the lower arithmetic cost which allow lower execution time
at minor degradation in performance,
which suggests a favorable trade-off.

As limitations of the present work,
we mention:
\begin{enumerate}[i)]
	\item the quasi-orthogonality property introduces error in the 3D inverse computation. This error is negligible for low bitrates.
	It tends, however, to be more prominent in terms of relative PSNR at high bitrates compared to the exact 3D~DHT; this increase is not observed for the SSIM metric, which suggests that the error is almost unnoticeable for the human vision;
	\item the proposed methods do not reduce the memory usage. This property is expected since the transforms are designed to have reduced arithmetic complexity,
	but they still present the same input and output sizes and the same transform lengths.
\end{enumerate}

For future work,
we suggest:
\begin{enumerate}[i)]
	\item implementing dedicate circuits in FPGA and ASICs for computing the proposed 3D~DHT in order to evaluate the components and area reduction;
	\item evaluating how well the proposed methods perform to compress natural videos and other 3D image data;
	\item expanding the current approach for higher lengths 3D~transforms, such as transforms of sizes $16 \times 16 \times 16$ and $32 \times 32 \times 32$.
	\item investigating the relation of the DHT and its approximate versions with the graph Fourier transform and possible applications in wireless sensor networks;
	\item investigating higher dimension DHT approximations such as 4D~DHT and 5D~DHT and possible applications at compressing multidimensional image data such as lightfield data.
\end{enumerate}

\section*{Acknowledgment}

	The authors would like to thank Conselho Nacional de Desenvolvimento Cient\'{\i}fico e Tecnol\'ogico (CNPq),
Brazil, for partially supporting this work.
The first author thanks to Crist\'ov\~ao Z. Rufino, M.Sc., for the valuable discussion he provided.

{\small
\singlespacing
\bibliographystyle{siam}
\bibliography{bib}

\begin{thebibliography}{10}

\bibitem{ahmed1975}
{\sc N.~Ahmed and K.~R. Rao}, {\em Orthogonal Transforms for Digital Signal
  Processing}, Springer, 1975.

\bibitem{arm_elf}
{\sc {ARM Developer}}, {\em {ARM ELF} specification}.

\bibitem{sytemv_abi}
{\sc {AT\&T, The Santa~CruzOperation~Inc.}}, {\em System {V} Application Binary
  Interface}, 4.1~ed., 1997.

\bibitem{barredatabase}
{\sc S.~Barr\'e}, {\em Medical imaging samples}.

\bibitem{bc2012}
{\sc F.~M. Bayer and R.~J. Cintra}, {\em {DCT}-like transform for image
  compression requires~14 additions only}, Electronics Letters, 48 (2012),
  pp.~919--921.

\bibitem{Bernstein2009}
{\sc D.~S. Bernstein}, {\em Matrix Mathematics: Theory, Facts, and Formulas},
  Princeton University Press, 2009.

\bibitem{bhaskaran1997}
{\sc V.~Bhaskaran and K.~Konstantinides}, {\em Image and Video Compression
  Standards}, Kluwer Academic Publishers, 1997.

\bibitem{Blahut2010}
{\sc R.~E. Blahut}, {\em Fast Algorithms for Signal Processing}, Cambridge
  University Press, 2010.

\bibitem{bas2008}
{\sc S.~Bouguezel, M.~O. Ahmad, and M.~N.~S. Swamy}, {\em Low-complexity
  8$\times$8 transform for image compression}, Electronics Letters, 44 (2008),
  pp.~1249--1250.

\bibitem{bas2013}
\leavevmode\vrule height 2pt depth -1.6pt width 23pt, {\em Binary discrete
  cosine and {H}artley transforms}, IEEE Transactions on Circuits and Systems
  I: Regular Papers, 60 (2013), pp.~989--1002.

\bibitem{boussakta2001radix}
{\sc S.~Boussakta, O.~H. Alshibami, and M.~Y. Aziz}, {\em Radix-$2 \times 2
  \times 2$ algorithm for the {3-D} discrete {H}artley transform}, IEEE
  Transactions on Signal Processing, 49 (2001), pp.~3145--3156.

\bibitem{bozinovic2003scan}
{\sc N.~Bozinovi\'c and J.~Konrad}, {\em Scan order and quantization for
  {3D-DCT} coding}, in Proceedings of {SPIE} Visual Communications and Image
  Processing, vol.~5150, 2003, p.~1205.

\bibitem{bozinovic2005}
\leavevmode\vrule height 2pt depth -1.6pt width 23pt, {\em Motion analysis in
  {3D DCT} domain and its application to video coding}, Signal Processing:
  Image Communication, 20 (2005), pp.~510--528.

\bibitem{bracewell1983discrete}
{\sc R.~N. Bracewell}, {\em Discrete {H}artley transform}, Journal of the
  Optical Society of America, 73 (1983), pp.~1832--1835.

\bibitem{bracewell1986fast}
{\sc R.~N. Bracewell, O.~Buneman, H.~Hao, and J.~Villasenor}, {\em Fast
  two-dimensional {H}artley transform}, Proceedings of the IEEE, 74 (1986),
  pp.~1282--1283.

\bibitem{britanak2007discrete}
{\sc V.~Britanak, P.~Yip, and K.~R. Rao}, {\em Discrete Cosine and Sine
  Transforms}, Academic Press, 2007.

\bibitem{chan1997variable}
{\sc Y.-L. Chan and W.-C. Siu}, {\em Variable temporal-length {3-D} discrete
  cosine transform coding}, IEEE Transactions on Image Processing, 6 (1997),
  pp.~758--763.

\bibitem{chien2018tensor}
{\sc J.-T. Chien and Y.-T. Bao}, {\em Tensor-factorized neural networks}, IEEE
  Transactions on Neural Networks and Learning Systems, 29 (2018),
  pp.~1998--2011.

\bibitem{chiper2013novel}
{\sc D.~F. Chiper}, {\em A novel {VLSI DHT} algorithm for a highly modular and
  parallel architecture}, IEEE Transactions on Circuits and Systems II: Express
  Briefs, 60 (2013), pp.~282--286.

\bibitem{cintra2011integer}
{\sc R.~J. Cintra}, {\em An integer approximation method for discrete
  sinusoidal transforms}, Circuits, Systems, and Signal Processing, 30 (2011),
  pp.~1481--1501.

\bibitem{cb2011}
{\sc R.~J. Cintra and F.~M. Bayer}, {\em A {DCT} approximation for image
  compression}, IEEE Signal Processing Letters, 18 (2011), pp.~579--582.

\bibitem{Cintra2018chapter}
{\sc R.~J. Cintra, F.~M. Bayer, Y.~Pauchard, and A.~Madanayake}, {\em
  Low-Complexity {DCT} Approximations for Biomedical Signal Processing in Big
  Data}, CRC Press, 1~ed., 2018, p.~624.

\bibitem{Cintra2014-sigpro}
{\sc R.~J. Cintra, F.~M. Bayer, and C.~J. Tablada}, {\em Low-complexity 8-point
  {DCT} approximations based on integer functions}, Signal Processing, 99
  (2014), pp.~201--214.

\bibitem{coutinho2017low}
{\sc V.~A. Coutinho, R.~J. Cintra, and F.~M. Bayer}, {\em Low-complexity
  multidimensional {DCT} approximations for high-order tensor data
  decorrelation}, IEEE Transactions on Image Processing, 26 (2017),
  pp.~2296--2310.

\bibitem{da2016orthogonal}
{\sc T.~L.~T. da~Silveira, F.~M. Bayer, R.~J. Cintra, S.~Kulasekera,
  A.~Madanayake, and A.~J. Kozakevicius}, {\em An orthogonal 16-point
  approximate {DCT} for image and video compression}, Multidimensional Systems
  and Signal Processing, 27 (2016), pp.~87--104.

\bibitem{lathauwer1998}
{\sc L.~De~Lathauwer and B.~De~Moor}, {\em From matrix to tensor: Multilinear
  algebra and signal processing}, in Institute of Mathematics and Its
  Applications Conference Series, vol.~67, Citeseer, 1998, pp.~1--16.

\bibitem{lathauwer2000best}
{\sc L.~De~Lathauwer, B.~De~Moor, and J.~Vandewalle}, {\em On the best rank-1
  and rank-({$R_1, R_2,\ldots, R_N$}) approximation of higher-order tensors},
  SIAM Journal on Matrix Analysis and Applications, 21 (2000), pp.~1324--1342.

\bibitem{descour2002toward}
{\sc M.~R. Descour, A.~Karkkainen, J.~D. Rogers, C.~Liang, R.~S. Weinstein,
  J.~T. Rantala, B.~Kilic, E.~Madenci, R.~R. Richards-Kortum, E.~V. Anslyn,
  et~al.}, {\em Toward the development of miniaturized imaging systems for
  detection of pre-cancer}, IEEE Journal of Quantum Electronics, 38 (2002),
  pp.~122--130.

\bibitem{dousty2016multifocus}
{\sc M.~Dousty, S.~Daneshvar, and R.~C. Sotero}, {\em Multifocus image fusion
  via the {H}artley transform}, in Canadian Conference on Electrical and
  Computer Engineering (CCECE), IEEE, 2016, pp.~1--5.

\bibitem{duhamel1990fast}
{\sc P.~Duhamel and M.~Vetterli}, {\em Fast {F}ourier transforms: a tutorial
  review and a state of the art}, Signal Processing, 19 (1990), pp.~259--299.

\bibitem{duleba1999hartley}
{\sc I.~Duleba}, {\em {H}artley transform in compression of medical ultrasonic
  images}, in Proceedings of the International Conference on Image Analysis and
  Processing, IEEE, 1999, pp.~722--727.

\bibitem{fw1992}
{\sc E.~Feig and S.~Winograd}, {\em Fast algorithms for the discrete cosine
  transform}, IEEE Transactions on Signal Processing, 40 (1992),
  pp.~2174--2193.

\bibitem{gao2011low}
{\sc Y.~Gao, Y.~Zheng, S.~Diao, W.-D. Toh, C.-W. Ang, M.~Je, and C.-H. Heng},
  {\em Low-power ultrawideband wireless telemetry transceiver for medical
  sensor applications}, IEEE Transactions on Biomedical Engineering, 58 (2011),
  pp.~768--772.

\bibitem{gerek20062}
{\sc O.~N. Gerek and A.~E. {\c{C}}etin}, {\em A {2-D} orientation-adaptive
  prediction filter in lifting structures for image coding}, IEEE Transactions
  on Image Processing, 15 (2006), pp.~106--111.

\bibitem{Gonzalez2001}
{\sc R.~C. Gonzalez and R.~Woods}, {\em Digital Image Processing}, Prentice
  Hall, 2001.

\bibitem{grigoryan2004novel}
{\sc A.~M. Grigoryan}, {\em A novel algorithm for computing the {1-D} discrete
  {H}artley transform}, IEEE Signal Processing Letters, 11 (2004),
  pp.~156--159.

\bibitem{guo2000efficient}
{\sc J.-I. Guo}, {\em An efficient design for one-dimensional discrete
  {H}artley transform using parallel additions}, IEEE Transactions on Signal
  Processing, 48 (2000), pp.~2806--2813.

\bibitem{hao1987three}
{\sc H.~Hao and R.~N. Bracewell}, {\em A three-dimensional {DFT} algorithm
  using the fast {H}artley transform}, Proceedings of the IEEE, 75 (1987),
  pp.~264--266.

\bibitem{haweel2001}
{\sc T.~I. Haweel}, {\em A new square wave transform based on the {DCT}},
  Signal Processing, 82 (2001), pp.~2309--2319.

\bibitem{Heideman1988}
{\sc M.~T. Heideman}, {\em Multiplicative Complexity, Convolution, and the
  {DFT}}, Signal Processing and Digital Filtering, Springer-Verlag, 1988.

\bibitem{hoppner2019achieve}
{\sc S.~H{\"o}ppner, H.~Eisenreich, D.~Walter, U.~Steeb, A.~S.~C. Dmello,
  R.~Sinkwitz, H.~Bauer, A.~Oefelein, F.~Schraut, J.~Schreiter, et~al.}, {\em
  How to achieve world-leading energy efficiency using {22FDX} with adaptive
  body biasing on an {ARM} cortex-{M4 IoT SoC}}, in ESSDERC 2019-49th European
  Solid-State Device Research Conference (ESSDERC), IEEE, 2019, pp.~66--69.

\bibitem{hossain2010medical}
{\sc M.~F. Hossain, M.~R. Alsharif, and K.~Yamashita}, {\em Medical image
  enhancement based on nonlinear technique and logarithmic transform
  coefficient histogram matching}, in IEEE/ICME International Conference on
  Complex Medical Engineering, IEEE, 2010, pp.~58--62.

\bibitem{hou1987fastdht}
{\sc H.~S. Hou}, {\em The fast {H}artley transform algorithm}, IEEE
  Transactions on Computers, 100 (1987), pp.~147--156.

\bibitem{jacob2015fpga}
{\sc J.~A. Jacob and N.~S. Kumar}, {\em {FPGA} implementation of optimal
  {3D}-integer {DCT} structure for video compression}, The Scientific World
  Journal, 2015 (2015).

\bibitem{jiang2010novel}
{\sc L.~Jiang, H.~Shu, J.~Wu, L.~Wang, and L.~Senhadji}, {\em A novel
  split-radix fast algorithm for {2-D} discrete {H}artley transform}, IEEE
  Transactions on Circuits and Systems I: Regular Papers, 57 (2010),
  pp.~911--924.

\bibitem{Kim1994floatingfixed}
{\sc S.~Kim and W.~Sung}, {\em A floating-point to fixed-point assembly program
  translator for the {TMS 320C25}}, {IEEE} Transactions on Circuits and Systems
  II: Analog and Digital Signal Processing, 41 (1994), pp.~730 -- 739.

\bibitem{kouadria2013low}
{\sc N.~Kouadria, N.~Doghmane, D.~Messadeg, and S.~Harize}, {\em Low complexity
  {DCT} for image compression in wireless visual sensor networks}, Electronics
  Letters, 49 (2013), pp.~1531--1532.

\bibitem{kulasekera2015multi}
{\sc S.~Kulasekera, A.~Madanayake, D.~Suarez, R.~J. Cintra, and F.~M. Bayer},
  {\em Multi-beam receiver apertures using multiplierless 8-point approximate
  {DFT}}, in Radar Conference (RadarCon), IEEE, 2015, pp.~1244--1249.

\bibitem{Lee1997csd}
{\sc M.-H. Lee}, {\em {CSD} filter design for {VLSI} implementation of {GA-VSB}
  receiver}, {IEEE} Transactions on Consumer Electronics, 43 (1997),
  pp.~197--206.

\bibitem{lengwehasatit2004scalable}
{\sc K.~Lengwehasatit and A.~Ortega}, {\em Scalable variable complexity
  approximate forward {DCT}}, IEEE Transactions on Circuits and Systems for
  Video Technology, 14 (2004), pp.~1236--1248.

\bibitem{li2007multiview}
{\sc L.~Li and Z.~Hou}, {\em Multiview video compression with {3D-DCT}}, in ITI
  5th International Conference on Information and Communications Technology,
  2007, pp.~59--61.

\bibitem{li2013visualtracking}
{\sc X.~Li, A.~Dick, C.~Shen, A.~van~den Hengel, and H.~Wang}, {\em Incremental
  learning of {3D-DCT} compact representations for robust visual tracking},
  {IEEE} Transactions on Pattern Analysis and Machine Intelligence, 35 (2013),
  pp.~863--881.

\bibitem{Liang2001}
{\sc J.~Liang and T.~D. Tran}, {\em Fast multiplierless approximation of the
  {DCT} with the lifting scheme}, IEEE Transactions on Signal Processing, 49
  (2001), pp.~3032--3044.

\bibitem{zenodomr}
{\sc W.~R. Lionheart}, {\em An {MRI DICOM} data set of the head of a normal
  male human aged 52}.

\bibitem{liu2010color}
{\sc Z.~Liu, J.~Dai, X.~Sun, and S.~Liu}, {\em Color image encryption by using
  the rotation of color vector in {H}artley transform domains}, Optics and
  Lasers in Engineering, 48 (2010), pp.~800--805.

\bibitem{Loeffler1989}
{\sc C.~Loeffler, A.~Ligtenberg, and G.~S. Moschytz}, {\em Practical fast
  \mbox{1-D~DCT} algorithms with 11 multiplications}, {ICASSP} International
  Conference on Acoustics, Speech, and Signal Processing, 2 (1989),
  pp.~988--991.

\bibitem{madanayake2015low}
{\sc A.~Madanayake, R.~J. Cintra, V.~Dimitrov, F.~M. Bayer, K.~A. Wahid,
  S.~Kulasekera, A.~Edirisuriya, U.~Potluri, S.~Madishetty, and N.~Rajapaksha},
  {\em Low-power {VLSI} architectures for {DCT/DWT}: precision vs approximation
  for {HD} video, biomedical, and smart antenna applications}, IEEE Circuits
  and Systems Magazine, 15 (2015), pp.~25--47.

\bibitem{mandal2013separable}
{\sc J.~K. Mandal and S.~K. Ghosal}, {\em Separable discrete {H}artley
  transform based invisible watermarking for color image authentication
  ({SDHTIWCIA})}, in Advances in Computing and Information Technology,
  Springer, 2013, pp.~767--776.

\bibitem{meher2006scalable}
{\sc P.~K. Meher, T.~Srikanthan, and J.~C. Patra}, {\em Scalable and modular
  memory-based systolic architectures for discrete {H}artley transform}, IEEE
  Transactions on Circuits and Systems I: Regular Papers, 53 (2006),
  pp.~1065--1077.

\bibitem{mulla2014image}
{\sc A.~Mulla, J.~Baviskar, A.~Baviskar, and C.~Warty}, {\em Image compression
  scheme based on zig-zag {3D-DCT} and {LDPC} coding}, in International
  Conference on Advances in Computing, Communications and Informatics (ICACCI),
  2014, pp.~2380--2384.

\bibitem{narendra2016hartley}
{\sc K.~C. Narendra and S.~Satyanarayana}, {\em {H}artley transform based
  correlation filters for face recognition}, in International Conference on
  Signal Processing and Communications (SPCOM), IEEE, 2016, pp.~1--5.

\bibitem{Nguyen2011}
{\sc B.~P. Nguyen, C.-K. Chui, S.-H. Ong, and S.~Chang}, {\em An efficient
  compression scheme for {4-D} medical images using hierarchical vector
  quantization and motion compensation}, Computers in Biology and Medicine, 41
  (2011), pp.~843--856.

\bibitem{oliveira2017low}
{\sc P.~A. Oliveira, R.~J. Cintra, F.~M. Bayer, S.~Kulasekera, and
  A.~Madanayake}, {\em Low-complexity image and video coding based on an
  approximate discrete {T}chebichef transform}, IEEE Transactions on Circuits
  and Systems for Video Technology, 27 (2017), pp.~1066--1076.

\bibitem{oliveira2017jpeg}
{\sc P.~A.~M. Oliveira, R.~S. Oliveira, R.~J. Cintra, F.~M. Bayer, and
  A.~Madanayake}, {\em {JPEG} quantisation requires bit-shifts only},
  Electronics Letters, 53 (2017), pp.~588--590.

\bibitem{onuki2017embedded}
{\sc T.~Onuki, W.~Uesugi, A.~Isobe, Y.~Ando, S.~Okamoto, K.~Kato, T.~R. Yew,
  J.~Wu, C.~C. Shuai, S.~H. Wu, et~al.}, {\em Embedded memory and {ARM}
  cortex-{M}0 core using 60-nm {C}-axis aligned crystalline
  indium--gallium--zinc oxide {FET} integrated with 65-nm {Si CMOS}}, IEEE
  Journal of Solid-State Circuits, 52 (2017), pp.~925--932.

\bibitem{oppenheim2010discrete}
{\sc A.~V. Oppenheim and R.~W. Schafer}, {\em Discrete-time Signal Processing},
  Pearson Higher Education, 3rd~ed., 2010.

\bibitem{osirixdatabase}
{\sc {OsiriX DICOM}}, {\em {O}siri{X} {DICOM} image sample sets}.

\bibitem{papitha2013compression}
{\sc J.~Papitha, G.~M. Nancy, and D.~Nedumaran}, {\em Compression techniques on
  {MR} image—a comparative study}, in International Conference on
  Communications and Signal Processing (ICCSP), IEEE, 2013, pp.~367--371.

\bibitem{Pauchard2015}
{\sc Y.~{Pauchard}, R.~J. {Cintra}, A.~{Madanayake}, and F.~M. {Bayer}}, {\em
  Fast computation of residual complexity image similarity metric using
  low-complexity transforms}, IET Image Processing, 9 (2015), pp.~699--708.

\bibitem{Rao2001}
{\sc K.~R. Rao and P.~Yip}, {\em The Transform and Data Compression Handbook},
  {CRC} Press {LLC}, 2001.

\bibitem{rao2014discrete}
\leavevmode\vrule height 2pt depth -1.6pt width 23pt, {\em Discrete Cosine
  Transform: Algorithms, Advantages, Applications}, Academic Press, San Diego,
  CA, 2014.

\bibitem{Rizkalla2002floatingdct}
{\sc M.~Rizkalla, P.~M.~Ei-Sharkawy~Salama, and B.~Dukel}, {\em Implementation
  of floating point fast discrete cosine transform}, The 45th Midwest Symposium
  on Circuits and Systems (MWSCAS), 2 (2002), pp.~II--17 -- II--20.

\bibitem{rubodatabase}
{\sc {Rubo Medical Imaging}}, {\em Rubo medical imaging dataset}.

\bibitem{sawant2011balanced}
{\sc S.~Sawant and D.~A. Adjeroh}, {\em Balanced multiple description coding
  for {3D~DCT} video}, IEEE Transactions on Broadcasting, 57 (2011),
  pp.~765--776.

\bibitem{Servais1997}
{\sc M.~Servais and G.~de~Jager}, {\em Video compression using the three
  dimensional discrete cosine transform {(3D-DCT)}}, in Proceedings of the 1997
  South African Symposium on Communications and Signal Processing~(COMSIG),
  IEEE, 1997, pp.~27--32.

\bibitem{song2013local}
{\sc T.~Song and H.~Li}, {\em Local polar {DCT} features for image
  description}, {IEEE} Signal Processing Letters, 20 (2013), pp.~59--62.

\bibitem{sorensen1985computing}
{\sc H.~Sorensen, D.~Jones, C.~Burrus, and M.~Heideman}, {\em On computing the
  discrete {H}artley transform}, IEEE Transactions on Acoustics, Speech, and
  Signal Processing, 33 (1985), pp.~1231--1238.

\bibitem{suarez2014multi}
{\sc D.~Suarez, R.~J. Cintra, F.~M. Bayer, A.~Sengupta, S.~Kulasekera, and
  A.~Madanayake}, {\em Multi-beam {RF} aperture using multiplierless {FFT}
  approximation}, Electronics Letters, 50 (2014), pp.~1788--1790.

\bibitem{sunder2005performance}
{\sc R.~S. Sunder, C.~Eswaran, and N.~Sriraam}, {\em Performance evaluation of
  {3-D} transforms for medical image compression}, in Proceedings of the
  International Conference on Electro Information Technology, IEEE, 2005,
  pp.~6--pp.

\bibitem{sunder2006medical}
{\sc R.~S. Sunder, C.~Eswaran, and N.~Sriraam}, {\em Medical image compression
  using {3-D} {H}artley transform}, Computers in Biology and Medicine, 36
  (2006), pp.~958--973.

\bibitem{rasp_pipico}
{\sc {The Raspberry Pi Foundation}}, {\em {R}aspberry {P}i {P}ico}.

\bibitem{rasp_pipico_sdk}
\leavevmode\vrule height 2pt depth -1.6pt width 23pt, {\em {R}aspberry {P}i
  {P}ico {S}oftware {D}evelopment {K}it ({SDK})}.

\bibitem{voronenko2009algebraic}
{\sc Y.~Voronenko and M.~Puschel}, {\em Algebraic signal processing theory:
  {C}ooley--{T}ukey type algorithms for real {DFT}s}, IEEE Transactions on
  Signal Processing, 57 (2009), pp.~205--222.

\bibitem{Wang2004}
{\sc Z.~Wang, A.~C. Bovik, H.~R. Sheikh, and E.~P. Simoncelli}, {\em Image
  quality assessment: from error visibility to structural similarity}, IEEE
  Transactions on Image Processing, 13 (2004), pp.~600--612.

\bibitem{watson1986separable}
{\sc A.~B. Watson and A.~Poirson}, {\em Separable two-dimensional discrete
  {H}artley transform}, Journal of the Optical Society of America {A}, 3
  (1986), pp.~2001--2004.

\bibitem{xue2015automatic}
{\sc M.~Xue, A.~Mian, W.~Liu, and L.~Li}, {\em Automatic {4D} facial expression
  recognition using {DCT} features}, in Winter Conference on Applications of
  Computer Vision, IEEE, 2015, pp.~199--206.

\bibitem{yakovlev2012implantable}
{\sc A.~Yakovlev, S.~Kim, and A.~Poon}, {\em Implantable biomedical devices:
  Wireless powering and communication}, IEEE Communications Magazine, 50
  (2012).

\bibitem{zaharia2002adaptive}
{\sc R.~Zaharia, A.~Aggoun, and M.~McCormick}, {\em Adaptive {3D-DCT}
  compression algorithm for continuous parallax {3D} integral imaging}, Signal
  Processing: Image Communication, 17 (2002), pp.~231--242.

\bibitem{zhang2018recryptor}
{\sc Y.~Zhang, L.~Xu, Q.~Dong, J.~Wang, D.~Blaauw, and D.~Sylvester}, {\em
  Recryptor: A reconfigurable cryptographic cortex-{M0} processor with
  in-memory and near-memory computing for {IoT} security}, IEEE Journal of
  Solid-State Circuits, 53 (2018), pp.~995--1005.

\end{thebibliography}
}

\end{document}